\documentclass[journal]{IEEEtran}
\ifCLASSINFOpdf
\else
   \usepackage{graphicx}
   \graphicspath{{../eps/}}
   \DeclareGraphicsExtensions{.eps}
\fi

\usepackage{amsmath}
\usepackage{amsfonts,amssymb}
\usepackage{amssymb}
\usepackage{wrapfig}
\usepackage{psfrag}
\usepackage{epstopdf}
\usepackage{cite}
\usepackage{graphicx}
\usepackage{subfigure}
\usepackage{threeparttable}
\usepackage{cases}
\usepackage{subeqnarray}
\usepackage{color}
\usepackage{underscore}
\usepackage{verbatim}
\usepackage{bm}
\usepackage{stfloats}
\usepackage{xpatch}
\usepackage{multicol}       
\usepackage{multirow}       
\usepackage{array}          

\newtheorem{algorithm}{Algorithm}
\newtheorem{proposition}{Proposition}

\usepackage{algorithm}
\usepackage{algorithmic}

\makeatletter
\def\changeBibColor#1{%
  \in@{#1}{}
  \ifin@\color{red}\else\normalcolor\fi
}
\xpatchcmd\@bibitem
  {\item}
  {\changeBibColor{#1}\item}
  {}{\fail}
\xpatchcmd\@lbibitem
  {\item}
  {\changeBibColor{#2}\item}
  {}{\fail}
\makeatother

\begin{document}
\title{Flexible XL-MIMO via Array Configuration Codebook: Codebook Design and Array Configuration Training}
%
%
%
\author{Haiquan~Lu,
        Hongqi~Min,
        Yong~Zeng,~\IEEEmembership{Fellow,~IEEE,}
        and
        Shaodan~Ma,~\IEEEmembership{Senior Member,~IEEE}
\thanks{Haiquan Lu is with the School of Electronic and Optical Engineering, Nanjing University of Science and Technology, Nanjing 210094, China, and is also with the State Key Laboratory of Internet of Things for Smart City and the Department of Electrical and Computer Engineering, University of Macau, Macao SAR, China. He was with the National Mobile Communications Research Laboratory, Southeast University, Nanjing 210096, China (e-mail: haiq_lu@163.com).}
\thanks{Hongqi Min and Yong Zeng are with the National Mobile Communications Research Laboratory, Southeast University, Nanjing 210096, China. Yong Zeng is also with the Purple Mountain Laboratories, Nanjing 211111, China (e-mail: \{minhq, yong_zeng\}@seu.edu.cn). (\emph{Corresponding author: Yong Zeng.})}
\thanks{Shaodan Ma is with the State Key Laboratory of Internet of Things for Smart City and the Department of Electrical and Computer Engineering, University of Macau, Macao SAR, China (e-mail: shaodanma@um.edu.mo).}
}

\maketitle

\begin{abstract}
 Extremely large-scale multiple-input multiple-output (XL-MIMO) emerges as a promising technology to achieve unprecedented enhancements in spectral efficiency and spatial resolution, via orders-of-magnitude increase in the antenna array size. However, the practical issues of high hardware cost and power consumption pose great challenges towards the cost-effective implementation of XL-MIMO. To address such challenges, this paper proposes a novel concept called array configuration codebook (ACC), which enables flexible XL-MIMO cost-effectively and improves the system performance compared with conventional antenna selection (AS) schemes with limited number of radio-frequency (RF) chains. Specifically, ACC refers to a set of pre-designed array configuration codewords, where each codeword specifies the positions of activated antenna pixels. Then, flexible XL-MIMO architecture can be enabled via dynamical pixel activation based on the designed ACC, without having to exhaustively try all possible combinations of the antenna pixels activations. As an illustration, we give a specific codebook design, encompassing the classic compact array (CA) where all activated pixels are separated by half wavelength, uniform sparse array (USA), modular array (MoA), nested array (NA), and co-prime array (CPA), and each codeword is specified by one array configuration parameter. With the designed ACC, array configuration training is considered for multi-user equipment (UE) communication to maximize the sum rate. To reduce the training overhead of exhaustive scanning, a two-stage scanning scheme is proposed, including the array- and pixel-level scanning. For comparison, the greedy AS scheme is proposed, where the resulting incremental signal-to-interference-plus-noise ratio (SINR) expression by activating antenna pixel sequentially is derived in closed-form. Subsequently, array configuration training is extended to the wireless localization scenario. Simulation results are provided to demonstrate the effectiveness of codeword optimization for scenarios of multi-UE communication and wireless localization.
\end{abstract}


\IEEEpeerreviewmaketitle
\section{Introduction}\label{sectionIntroduction}
 Over the past few decades, wireless communication networks have undergone transformative evolutions from the first-generation (1G) to fifth-generation (5G), where network capabilities and usage scenarios have been significantly enhanced and diversified. As 5G achieves widespread commercialization, various visions and extensive research efforts on the future 6G are in full swing. Notably, the International Telecommunication Union-Radiocommunication Sector (ITU-R) has identified six major usage scenarios and fifteen capabilities of 6G in 2023 \cite{ITU}, marking a critical milestone of 6G development. However, current 5G technologies are challenging to fulfill the ambitious performance requirements of 6G, e.g., the peak data rate increases to terabit-per-second (Tbps), latency is reduced to sub-millisecond level, and connection density reaches $10^8$ devices/km$^2$. To satisfy these requirements, many promising technologies have been proposed, such as extremely large-scale multiple-input multiple-output (XL-MIMO) \cite{bjornson2019massive,lu2022communicating,lu2024tutorial,han2023toward,wang2024extremely}, Terahertz (THz) communication \cite{akyildiz2014teranets}, intelligent reflecting surface/reconfigurable intelligent surface (IRS/RIS) \cite{wu2021intelligent,di2020smart}, integrated sensing and communication (ISAC) \cite{liu2022integrated,dai2025tutorial}, and channel knowledge map (CKM) \cite{zeng2021toward}. In particular, XL-MIMO is a natural evolution of existing massive MIMO, where the antenna number at the base station (BS) is boosted by at least an order of magnitude, reaching hundreds or even thousands of antennas, so as to pursue a substantial improvement in spectral efficiency and spatial resolution  \cite{bjornson2019massive,lu2024tutorial}. In this context, many research endeavors have been devoted to XL-MIMO technique, including the near-field channel modeling, performance analysis, and practical design issues \cite{lu2024tutorial,han2023toward,wang2024extremely}.

 Despite the significant performance enhancement, the implementation of XL-MIMO faces practical issues of high hardware cost and power consumption. For example, for fully digital beamforming, each antenna includes one dedicated radio-frequency (RF) chain and antenna pixel. While the antenna pixel, e.g., patch antenna, is cost-effective, RF chain is expensive and power-hungry, particularly due to the components of power amplifier, digital-to-analog and analog-to-digital converter (DAC/ADC) \cite{yang2018digital}. To tackle this issue, hybrid beamforming architecture is widely regarded as an effective method, where the digital beamforming is performed with a limited number of RF chains and analog part is implemented via phase shifters or switches \cite{el2014spatially,heath2016overview,sohrabi2017hybrid}. Compared to the fully digital beamforming, hybrid beamforming strikes a balance between the cost/power consumption and system performance. Alternatively, lens antenna array is another cost-effective implementation method, which consists of an electromagnetic lens and matched antenna array \cite{zeng2016millimeter}. In particular, the lens is capable of providing variable phase shifts across different regions of its continuous aperture, thus enabling a novel analog beamforming realization.

 Besides the hybrid beamforming architecture, antenna selection (AS) also can reduce the number of RF chains, where only a subset of antennas with favorable channel conditions are activated, without incurring much channel capacity loss \cite{sanayei2004antenna,gharavi2004fast,lu2025fast}. Sparse array is another promising approach for improving performance with a limited number of antennas. Specifically, conventional BS is typically equipped with a compact array (CA), where neighbouring antennas are separated by half wavelength to avoid causing the mutual coupling effect and grating lobes. Meanwhile, half wavelength roughly corresponds to the channel coherence distance where each antenna experiences independent fading in rich scattering environment, thus improving the spatial diversity gain \cite{li2025sparse}. By contrast, sparse array adjusts the antenna spacing to enable an enlarged array aperture than the classic CA when equipped with the identical number of antennas. Many research efforts have been devoted to developing various sparse array architectures, e.g., uniform sparse array (USA) and non-uniform sparse array (NUSA), where NUSA includes the modular array (MoA), nested array (NA), co-prime array (CPA), minimum redundancy array (MRA), minimum holes array (MHA) \cite{liXiang2025sparse} and so on, aiming at enhancing the spatial resolution for communication \cite{wang2023can,wang2024enhancing,li2024multi,lu2024group}, as well as increasing the virtual aperture and sensing degree of freedom (DoF) in radar sensing \cite{pal2010nested,vaidyanathan2011sparse,li2025sparse,min2025integrated}. In particular, unlike CA, USA, MoA, NA, and CPA, there are no closed-form expressions of array architectures for both MRA and MHA, and thus exhaustive search is required to determine the corresponding array architectures \cite{liXiang2025sparse}.
 
 In addition, fluid antenna system (FAS) \cite{wong2021fluid} or movable antenna (MA) \cite{zhu2024movable,lu2024group,zhu2025tutorial} emerges as a promising technology for performance enhancement. By endowing each antenna the capability of flexible movement and/or rotation through the mechanical or equivalent electronic control, MA is able to break the limitation of fixed-position antennas (FPAs) in the current BS, which helps to fully exploit the spatial DoF for enhanced multiplexing and diversity performance. More recently, ray antenna array (RAA) was proposed to achieve enhanced beamforming gain and uniform angular resolution \cite{dong2025ray,jiang2025ray,dong2025novel}. Specifically, RAA employs a ray-like structure antenna pixel placement, where cost-effective antenna pixels along each ray are directly connected and constitute a simple uniform linear array (sULA) with a pre-designed orientation. This enables each ray to form a beam pointing towards its physical orientation, without resorting to any analog or digital beamforming.

 While the aforementioned technologies effectively reduce the number of RF chains, they also face many practical implementation issues. For example, the number of required phase shifters in the fully-connected hybrid beamforming architecture equals the number of antennas multiplying the number of RF chains, and their costs will become unaffordable in XL-MIMO with hundreds or even thousands of antennas, particularly for high frequency bands such as millimeter wave (mmWave) and THz. Regarding AS, it suffers from the beamforming gain loss since only a small subset of antennas are activated. Besides, when the number of candidate antennas drastically increases, the implementation complexity to exhaustively search all possible antenna combinations is prohibitive for AS. Meanwhile, the pilot overhead and channel estimation complexity are exacerbated by the high-dimensional XL-MIMO channel matrices. Moreover, MA involves the issues of complicated position optimization and movement control, as well as mechanical movement-induced delay \cite{lu2024group,zhu2025tutorial}.

 \begin{figure}[!t]
 \centering
 \centerline{\includegraphics[width=2.85in,height=1.7in]{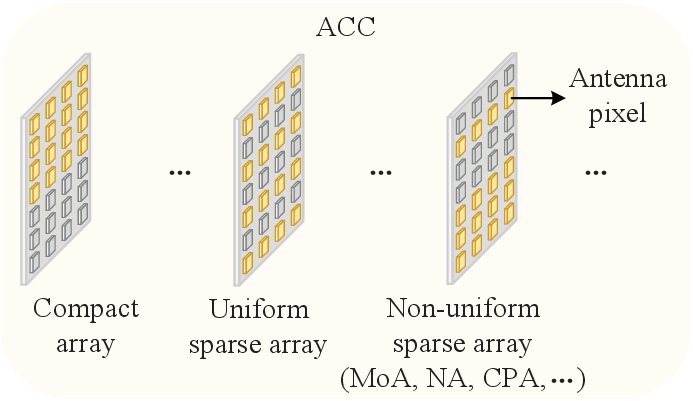}}
 \caption{An illustration of ACC enabled flexible XL-MIMO, where the yellow and gray elements correspond to the activated and deactivated antenna pixels, respectively. The flexible XL-MIMO architecture can be enabled via dynamic pixel activation/selection.}
 \label{fig:ACCIllustration}
 \end{figure}
 Motivated by the above discussions, in this paper, we propose a novel concept called array configuration codebook (ACC), which practically enables flexible XL-MIMO architecture efficiently. Specifically, ACC is a set of pre-designed array configuration codewords, where each codeword specifies the positions of activated antenna pixels.
 It is worth mentioning that codebook-based wireless communications have been widely adopted, such as source-channel coding codebook \cite{gunduz2024joint} and beamforming codebook \cite{wang2009beam}. Moreover, flexible XL-MIMO consists of a massive number of cost-effective antenna pixels, and a subset of antenna pixels are dynamically activated based on the designed ACC, as illustrated in Fig.~\ref{fig:ACCIllustration}, and connected to the limited number RF chains. Compared to the conventional AS, the ACC based pixel activation significantly reduces the number of possible antenna combinations, since a large number of ineffective array configurations are excluded for consideration. This also reduces the implementation complexity of pixel selection network suffered by AS. The main contributions of this paper are summarized as follows:

 \begin{itemize}[\IEEEsetlabelwidth{12)}]
 \item First, we propose a novel concept called ACC to enable flexible XL-MIMO. As an illustration, we present a specific ACC design example, which is composed of the classic CA, USA, MoA, NA, and CPA architectures. For each architecture, the codebook is designed based on its array configuration parameter. Specifically, the reference position is the common array configuration parameter for all architectures, while the sparsity level, inter-module spacing level, the number of inner array's antenna pixels, and the number of the first array's antenna pixels are architecture-specific parameters for USA, MoA, NA, and CPA, respectively. In particular, the physical and virtual apertures of the array vary with array configuration parameters, which in turn impact the system performance.
 \item Second, to illustrate the benefits of ACC, the multi-user equipment (UE) communication scenario is considered. With the designed ACC, we formulate an array configuration training problem to maximize the sum rate of all UEs. The scheme of exhaustive scanning over ACC is first proposed to find the optimal array configuration codeword. To reduce the training overhead of exhaustive scanning, we propose an efficient two-stage scanning scheme, including the coarse array-level scanning and fine pixel-level scanning. Moreover, for comparison, a benchmark scheme of greedy AS is proposed, where one antenna pixel is selected to maximize the resulting sum rate in each step. In particular, the resulting incremental signal-to-interference-plus-noise ratio (SINR) expression by activating antenna pixel sequentially is derived in closed-form, i.e., the SINR of the previous step plus the increment of current antenna pixel, thus significantly reducing the computational complexity.
 \item Last, the multi-UE communication scenario is extended to wireless localization scenario, where array configuration training aims to minimize the root mean squared error (RMSE) of angle of arrival (AoA) estimation, and the two-stage scanning scheme can be similarly applied. Simulation results are presented to demonstrate the importance of codeword optimization, and necessity of considering different array architectures for different scenarios, due to their notable performance discrepancies.
 \end{itemize}

 The remainder of this paper is organized as follows. Section~\ref{sectionSystemModel} presents the system model and formulates a general optimization problem to maximize any utility function. Section~\ref{sectionACCDesign} gives a specific ACC design example composed of CA, USA, MoA, NA, and CPA. Section~\ref{sectionArrayTrainingCommunication} proposes an efficient two-stage array configuration training scheme for multi-UE communication, and a benchmark scheme of greedy AS. Section~\ref{sectionArrayTrainingLocalization} studies the array configuration training for wireless localization. Simulation results are presented in Section~\ref{sectionNumericalResults}. Finally, Section~\ref{sectionConclusion} concludes this paper.

 \emph{Notations:} Scalars are denoted by italic letters. Vectors and matrices are denoted by bold-face lower- and upper-case letters, respectively. ${{\mathbb{C}}^{M \times N}}$ and ${{\mathbb{R}}^{M \times N}}$ represent the space of $M \times N$ complex-valued and real-valued matrices, respectively. ${{\bf{I}}_N}$ denotes an $N \times N$ identity matrix. For a vector ${\bf{x}}$, $\left\| {\bf{x}} \right\|$ denotes its Euclidean norm. For a matrix ${\bf A}$, its complex conjugate, transpose, and Hermitian transpose are denoted by ${\bf A}^*$, ${\bf A}^T$, ${\bf A}^H$, respectively. The distribution of a circularly symmetric complex Gaussian random vector with mean $\bf{x}$ and covariance matrix $\bf{\Sigma}$ is denoted by ${\cal CN}\left( {\bf{x},\bf{{\Sigma}}} \right)$; and $\sim$ stands for ``distributed as". The symbol ${\rm j}$ denotes the imaginary unit of complex numbers, with ${{\rm j}^2} =  - 1$. For real number $x$, $\left\lfloor x \right\rfloor $ and $\left\lceil x \right\rceil $ denote the floor and ceiling operations, respectively. ${\cal O}\left({\cdot}\right)$ denotes the standard big-O notation. $\left| {\cal W} \right|$ represents the cardinality of a set $\cal W$. $\odot$ and $\otimes$ denote the Khatri-Rao and Kronecker products, respectively. ${\rm{vec}}\left(\cdot \right)$ denotes the column-wise vectorization operator.

\section{System Model And Problem Formulation}\label{sectionSystemModel}
 \begin{figure}[!t]
 \centering
 \centerline{\includegraphics[width=3.5in,height=1.3in]{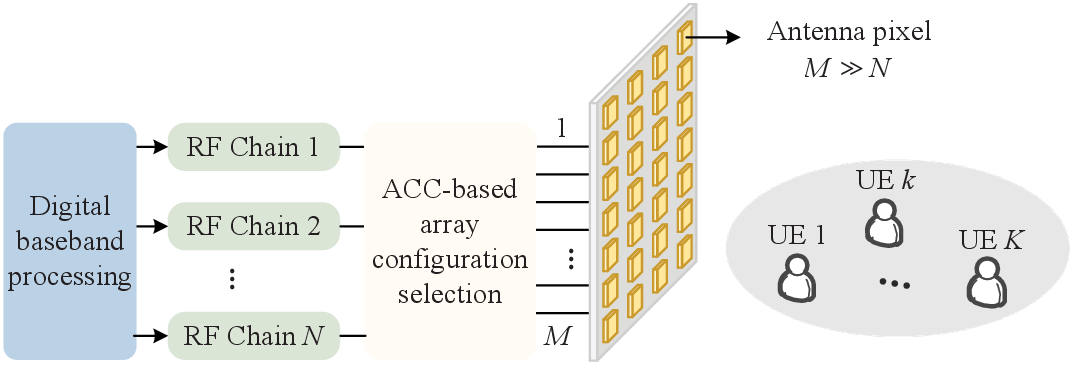}}
 \caption{A wireless communication, localization or sensing system with ACC-based flexible XL-MIMO.}
 \label{fig:systemModel}
 \end{figure}
 As shown in Fig.~\ref{fig:systemModel}, we propose an ACC-based wireless communication, localization or sensing system with a BS and $K$ UEs. The BS is equipped with an extremely large-scale array (XL-array) of $M \gg 1$ antenna pixels, with the position of pixel $m$ denoted as ${{\bf{s}}_m}$, and the number of RF chains is $N \ll M$. The array architecture selection module activates $N$ antenna pixels from $M$ pixels based on the ACC. Let ${{\bf{q}}_k}$ denote the location of the single-antenna UE $k$, and ${\bf{q}}_{k,l}$ denote the location of scatterer $l$ for UE $k$, $1 \le l \le {L_k}$, with ${L_k}$ being the number of scatterers experienced by UE $k$. In particular, for line-of-sight (LoS) channel of UE $k$, the scatterer corresponds directly to UE $k$ itself.

 Let ${\bf{\tilde w}} \triangleq {\left[ {{{\bf{s}}_1^T}, \cdots ,{{\bf{s}}_m^T}, \cdots ,{{\bf{s}}_M^T}} \right]^T}$. For given UE/scatterer location ${\bf{q}}$, the array response vector is a function of ${\bf{\tilde w}}$ and ${\bf{q}}$, denoted as ${\bf{a}}\left( {{\bf{\tilde w}},{\bf{q}}} \right) \in {\mathbb C}^{M \times 1}$. Then, the channel between the BS and UE $k$ is
 \begin{equation}\label{ChannelUEk}
 {{\bf{h}}_k}\left( {\bf{\tilde w}} \right) = \sum\limits_{l = 1}^{{L_k}} {{\alpha _{k,l}}{\bf{a}}\left( {{\bf{\tilde w}},{{\bf{q}}_{k,l}}} \right)},
 \end{equation}
 where ${\alpha _{k,l}}$ represents the complex-valued gain of multi-path $l$ for UE $k$. To reveal the essential idea of ACC, the deterministic channel model is considered, and the extension to stochastic channel model deserves further investigation in the future.
 
 For XL-MIMO systems, the number of antenna pixels $M$ is generally much larger than that of RF chains $N$, i.e., $M \gg N$. Then, dynamic pixel activation is applied, where $N$ antenna pixels are selected out of $M$ pixels. Let ${\cal W}$ denote the set of all possible antenna pixel position combinations. For example, when $M =4$ and $N=2$, the set ${\cal W}$ is
 \begin{equation}\label{antennaPixelSet}
 {\cal W} = \left\{ {\left[ \begin{array}{l}
{{\bf{s}}_1}\\
{{\bf{s}}_2}
\end{array} \right],\left[ \begin{array}{l}
{{\bf{s}}_1}\\
{{\bf{s}}_3}
\end{array} \right],\left[ \begin{array}{l}
{{\bf{s}}_1}\\
{{\bf{s}}_4}
\end{array} \right],\left[ \begin{array}{l}
{{\bf{s}}_2}\\
{{\bf{s}}_3}
\end{array} \right],\left[ \begin{array}{l}
{{\bf{s}}_2}\\
{{\bf{s}}_4}
\end{array} \right],\left[ \begin{array}{l}
{{\bf{s}}_3}\\
{{\bf{s}}_4}
\end{array} \right]} \right\},
 \end{equation}
 where each element of $\cal W$ is composed of $N$ activated antenna pixel positions and specifies an antenna pixel combination, with $\left| {\cal W} \right| = \binom{M}{N}$.

 Moreover, let ${\bf w} \in {\cal W}$ denote the vector of activated antenna pixel positions. The corresponding channel of UE $k$ after dynamic pixel activation is ${{\bf{h}}_k}( {{\bf w}}  ) \in {{\mathbb C}^{N \times 1}}$. Denote by $U( {{ {\bf w}},{\left\{ {{{\bf{v}}_k}} \right\}_{k = 1}^K} } )$ a utility function for communication, localization or sensing, where ${{\bf{v}}_k} \in {{\mathbb C}^{N \times 1}}$ is the beamforming with respect to (w.r.t.) UE $k$. This thus yields the following optimization problem
 \begin{equation}\label{optimizationProblem}
 \begin{aligned}
 \mathop {\max }\limits_{{{\bf w}},  {\left\{ {{{\bf{v}}_k}} \right\}_{k = 1}^K}  }&\ \  U\left( {{{\bf w}},{\left\{ {{{\bf{v}}_k}} \right\}_{k = 1}^K} } \right)\\
 {\rm{s.t.}}&\ \  { {\bf w}} \in {\cal W}.
 \end{aligned}
 \end{equation}

 Note that antenna positions in problem \eqref{optimizationProblem} can be optimally obtained by exhaustively searching all $\binom{M}{N}$ possible combinations. Though this approach is valid when $N$ and $M$ is small, it will become computationally prohibitive for massive MIMO or XL-MIMO systems. For example, when $N = 32$ antenna pixels are selected out of $M = 256$ pixels, the number of possible combinations is $\binom{256}{32} \approx 5.824 \times {10^{40}}$. To tackle this issue, we propose a new ACC based pixel activation scheme, as shown in the following section.

\section{Array Configuration Codebook}\label{sectionACCDesign}
 In this section, we propose the concept of ACC and give a specific design example.

 ACC refers to a set of pre-designed array configuration codewords, denoted as ${\cal W}^{\rm ACC}$, where each codeword specifies the positions of activated antenna pixels and form one array architecture. In contrast to exhaustive enumeration of all possible antenna combinations, ACC excludes a large number of ineffective array configurations from consideration, and is thus a subset of $\cal W$, i.e., ${\cal W}^{\rm ACC} \subset {\cal W}$. Moreover, the cardinality of ${\cal W}^{\rm ACC}$ is much smaller than that of ${\cal W}$, i.e., $\left| {{\cal W}^{\rm{ACC}}} \right| \ll \left| {\cal W} \right|$. As a result, problem \eqref{optimizationProblem} becomes
 \begin{equation}\label{optimizationProblemACC}
 \begin{aligned}
 \mathop {\max }\limits_{{\bf w},  {\left\{ {{{\bf{v}}_k}} \right\}_{k = 1}^K}  }&\ \  U\left( {{\bf w},{\left\{ {{{\bf{v}}_k}} \right\}_{k = 1}^K} } \right)\\
 {\rm{s.t.}}&\ \  {\bf w} \in {\cal W}^{\rm ACC},
 \end{aligned}
 \end{equation}
 which can significantly reduce the search process if ACC is properly designed. Note that when the underlying array architecture is unknown or irregular, the proposed ACC-based method can be still applied, albeit with potential performance loss.

\begin{figure*}[!t]
 \centering
 \centerline{\includegraphics[width=5.3in,height=2.7in]{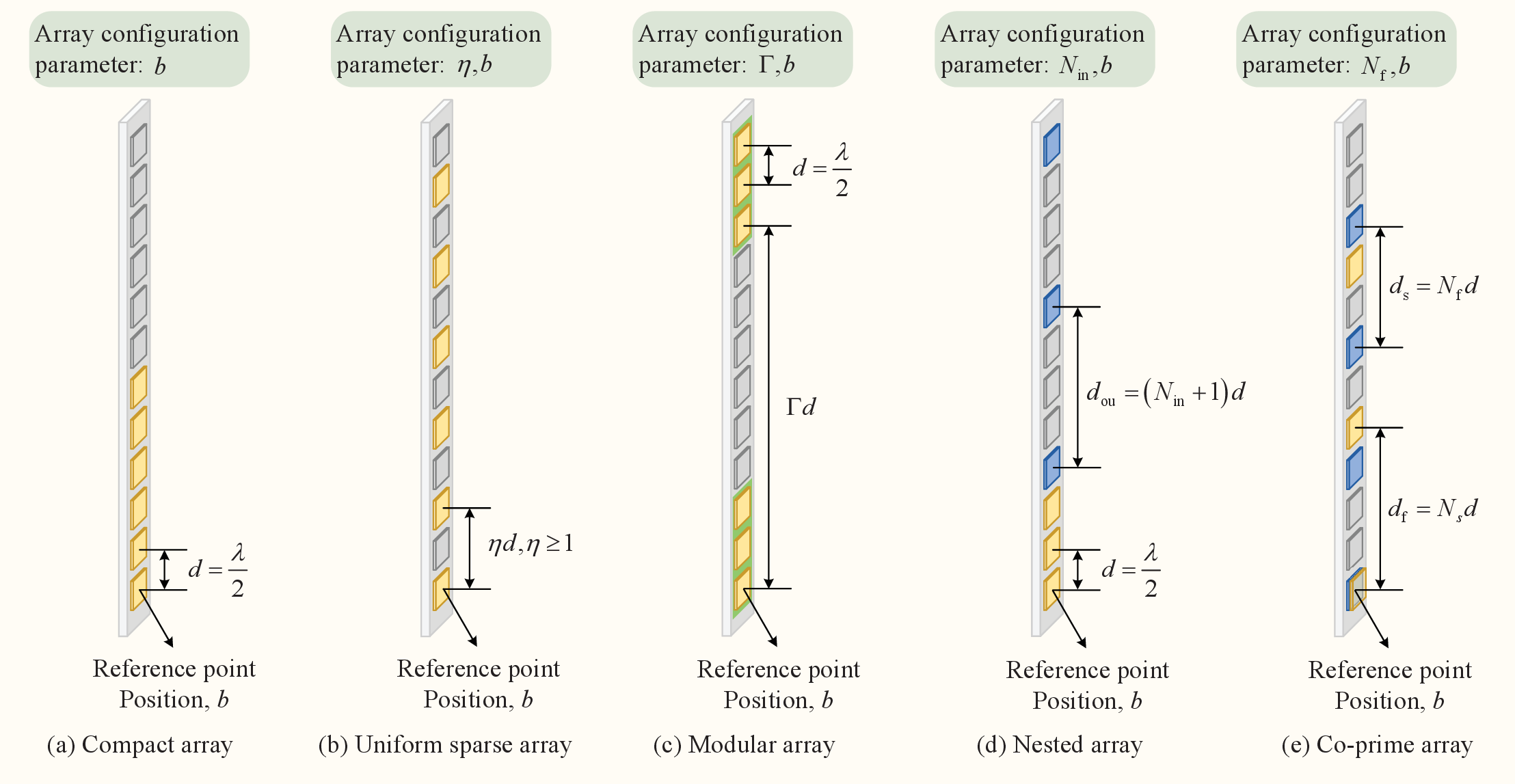}}
 \caption{An illustration of different array architectures, where the gray elements correspond to the deactivated pixels, and yellow/blue elements correspond to the activated pixels.}
 \label{fig:arrayArchitectureIllustration}
 \end{figure*}
 
 Next, we give one specific example of ACC design, which is composed of CA, USA, MoA, NA, and CPA, as illustrated in Fig.~\ref{fig:arrayArchitectureIllustration}. For ease of exposition, we consider an extremely large-scale uniform linear array (XL-ULA) with $M \gg 1$ antenna pixels. Without loss of generality, the XL-ULA is placed along the $y$-axis, with its center being the origin. The position of antenna pixel $m$ is ${{\bf{s}}_m} = {\left[ {0,{\delta _m}d} \right]^T}$, where ${\delta _m} = \left( {2m - M - 1} \right)/2$ and $d$ denotes the spacing between adjacent antenna pixels. As discussed in Section~\ref{sectionIntroduction}, to circumvent the issues of mutual coupling and grating lobes while reaping the spatial diversity gain, the spacing between adjacent antenna pixels is set to $d = {\lambda}/2$, with $\lambda$ being the signal wavelength.

\subsubsection{Compact Array}
 CA is a standard array architecture, where adjacent antenna pixels are typically separated by half wavelength, i.e., ${d} = \lambda /2$, as illustrated in Fig.~\ref{fig:arrayArchitectureIllustration}(a). Since all antenna pixels have the same $x$-axis coordinate value, the $y$-axis coordinate value is used to denote the pixel position. Without loss of generality, let the bottom pixel be the reference point, with its position denoted as ${s_{{\rm{ref}}}} = b$. The position of antenna pixel $n$ for CA is ${s_n} = {s_{{\rm{ref}}}} + \left( {n - 1} \right)d = b + \left( {n - 1} \right)d$, $1 \le n \le N$, and the physical dimension is ${D_{{\rm{CA}}}} = \left( {N - 1} \right){d}$. By varying the array configuration parameter of reference point position, we have ${b_{\min }} =  - \frac{{M - 1}}{2}d$ and ${b_{\max }} = \frac{{M - 1}}{2}d - \left( {N - 1} \right)d = \frac{{M - 2N + 1}}{2}d$. Thus, the total realizations of CA are ${B_{{\rm{CA}}}} = \left( {{b_{\max }} - {b_{\min }}} \right)/d + 1= M - N +1 $. The codebook for CA is
 \begin{equation}\label{codebookCA}
 {\cal W}_{{\rm{CA}}}  = \left\{ {{{\bf{w}}_{{\rm{CA}}}}\left( {{b_{\min }}} \right), \cdots ,{{\bf{w}}_{{\rm{CA}}}}\left( {{b_{\max }}} \right)} \right\},
 \end{equation}
 where the codeword ${{\bf{w}}_{{\rm{CA}}}}\left( b \right) \in {{\mathbb R}^{N \times 1}} = \left[ {b,b + d, \cdots } \right.$ ${\left. {b + \left( {N - 1} \right)d} \right]^T}$ consists of antenna pixel positions of the $b $-th realization for CA.

\subsubsection{Uniform Sparse Array}
 For USA, the spacing between adjacent antenna pixels is $\eta d$, where $\eta \ge 1$ represents the sparsity level \cite{wang2023can,lu2024group}, as illustrated in Fig.~\ref{fig:arrayArchitectureIllustration}(b). In particular, when $\eta = 1$, USA reduces to the classic CA. Similarly, let the bottom pixel be the reference point, with its position denoted as ${s_{{\rm{ref}}}} = b$. The position of antenna pixel $n$ for USA is ${s_n} = {s_{{\rm{ref}}}} + \left( {n - 1} \right)\eta d = b + \left( {n - 1} \right)\eta d$, and the physical dimension is ${D_{{\rm{USA}}}} \left(\eta\right) = \left( {N - 1} \right)\eta d$. Note that different from CA, the array configuration parameter for USA includes the sparsity level $\eta$ and reference point position $b$. Specifically, for any given sparsity level $\eta$, the minimum and maximum reference point positions are
 ${b_{\min }}\left({\eta}\right) =  - \frac{{M - 1}}{2}d$ and ${b_{\max }}\left({\eta}\right) = \frac{{M - 1}}{2}d - \left( {N - 1} \right)\eta d = \frac{{M - 1 - 2\left( {N - 1} \right)\eta }}{2}d$, respectively, where the maximum position is derived from the fact that the $N$-th pixel of USA cannot exceed the last pixel of the original array, i.e., $b + \left( {N - 1} \right)\eta d \le \frac{{M - 1}}{2}d$. Thus, for given $\eta$, the realizations of USA are $\left( {{b_{\max }}\left({\eta}\right) - {b_{\min }}\left({\eta}\right)} \right)/d + 1 = M - \left( {N - 1} \right)\eta $. Besides, the maximum sparsity level is ${\eta _{\max }} = \left\lfloor {\frac{{M - 1}}{{N - 1}}} \right\rfloor$, which is due to the fact that the dimension of USA is no greater than that of the original array, i.e., $\left( {N - 1} \right)\eta d \le \left( {M - 1} \right)d$. By combining all the sparsity levels, the total realizations of USA are ${B_{{\rm{USA}}}} = \sum\nolimits_{\eta  = 2}^{{\eta _{\max }}} \left({M - \left( {N - 1} \right)\eta }\right)  = \left( {{\eta _{\max }} - 1} \right)M - \left( {N - 1} \right)\left( {{\eta _{\max }} + 2} \right)\left( {{\eta _{\max }} - 1} \right)/2$, where the sparsity level starts from $\eta = 2$ so as to exclude the special case of CA. The codebook for USA is
 \begin{equation}\label{codebookUSA}
 {{\cal W}_{{\rm{USA}}}} = {\left\{ {{{\bf{w}}_{{\rm{USA}}}}\left( {\eta ,b} \right)} \right\}_{2 \le \eta  \le {\eta _{\max }},{b_{\min }}\left( \eta  \right) \le b \le {b_{\max }}\left( \eta  \right)}},
 \end{equation}
 where ${{\bf{w}}_{\rm{USA}}}\left( {\eta ,b} \right) \in {{\mathbb R}^{N \times 1}} = \left[ {b,b + \eta d, \cdots ,b + } \right.$ ${\left. {\left( {N - 1} \right)\eta d} \right]^T}$.

\subsubsection{Modular Array}
 As illustrated in Fig.~\ref{fig:arrayArchitectureIllustration}(c), MoA is composed of several modules, where adjacent antenna pixels within each module are separated by half-wavelength as in the classic CA, but the spacing between neighbouring modules is typically much larger than signal wavelength to accommodate practical deployment environment \cite{li2024multi,jeon2021mimo}. Specifically, we consider a MoA with fixed number of $Z$ modules, and each module has $N_{m} = N /Z$ antenna pixels. For notational convenience, we assume that $N /Z$ is an integer. Then, the module index $z$ and antenna pixel index $n$ belong to the integer sets ${\cal Z} = \left\{ {1, \cdots ,Z} \right\}$ and ${\cal N} = \left\{ {1, \cdots ,{N_m}} \right\}$, respectively. Besides, the inter-module spacing is $\Gamma d$, i.e., the separation between the first pixels of adjacent modules, with $\Gamma \ge {N_m}$. In particular, when $\Gamma = N_m$, MoA reduces to the classic CA. Let the bottom pixel of MoA be the reference point, with its position denoted as $s_{\rm{ref}} = b$. The position of antenna pixel $n$ for module $z$ is ${s_{z,n}} = b + \left( {z - 1} \right)\Gamma d + \left( {n - 1} \right)d$, $z \in {\cal Z}$, $n \in {\cal N}$. The physical dimension of MoA is $D_{\rm MoA} \left(\Gamma \right) = \left( {Z - 1} \right)\Gamma d + \left( {{N_m} - 1} \right)d$.

 The array configuration parameter for MoA is the inter-module spacing level $\Gamma$ and reference point position $b$. For given $\Gamma$, the minimum and maximum reference point positions are ${b_{\min }}\left( \Gamma \right) = - \frac{{M - 1}}{2}d$ and ${b_{\max }}\left( \Gamma \right) = \frac{{M - 2\left( {Z - 1} \right)\Gamma - 2{N_m} + 1}}{2}d$, respectively, where the maximum reference point position is derived from $b + \left( {Z - 1} \right)\Gamma d + \left( {{N_m} - 1} \right)d \le \frac{{M - 1}}{2}d$. Thus, the realizations of the MoA for given $\Gamma$ are $\left( {{b_{\max }}\left( \Gamma \right) - {b_{\min }}\left( \Gamma \right)} \right)/d + 1 = M - {N_m} + 1 - \left( {Z - 1} \right)\Gamma$. Besides, the maximum inter-module spacing level is ${\Gamma_{\max }} = \left\lfloor {\frac{{M - {N_m}}}{{Z - 1}}} \right\rfloor $, since the dimension of the MoA is no greater than that of the original array, i.e., $\left( {Z - 1} \right)\Gamma d + \left( {{N_m} - 1} \right)d \le \left( {M - 1} \right)d$. By combining all the inter-module spacing levels, the total realizations of MoA are ${B_{{\rm{MoA}}}} = \sum\nolimits_{\Gamma = {N_m} + 1}^{{\Gamma_{\max }}} {\left( {M - {N_m} + 1 - \left( {Z - 1} \right)\Gamma} \right)} =
 \left( {{\Gamma_{\max }} - {N_m}} \right)\left( {M - {N_m} + 1} \right) - \left( {Z - 1} \right)\left( {{N_m} + 1 + {\Gamma_{\max }}} \right) \times $ $\left( {{\Gamma_{\max }} - {N_m}} \right)/2$, where the inter-module spacing level starts from ${N_m} + 1$ to avoid the duplicate consideration of CA. The codebook for MoA is
 \begin{equation}\label{codebookMoA}
 {{\cal W}_{{\rm{MoA}}}} = {\left\{ {{{\bf{w}}_{{\rm{MoA}}}}\left( {\Gamma,b} \right)} \right\}_{{N_m} + 1 \le \Gamma \le {\Gamma_{\max }},{b_{\min }}\left( \Gamma \right) \le b \le {b_{\max }}\left( \Gamma \right)}},
 \end{equation}
 where ${{\bf{w}}_{{\rm{MoA}}}}\left( {\Gamma,b} \right) \in {{\mathbb R}^{N \times 1}} = \left[ {b, \cdots ,b + \left( {z - 1} \right)\Gamma d} \right.$ $ + \left( {n - 1} \right)d{\left. {, \cdots ,b + \left( {Z - 1} \right)\Gamma d + \left( {{N_m} - 1} \right)d} \right]^T}$.

\subsubsection{Nested Array}
 We consider a two-level NA, which is a non-uniform array composed of 
 an inner array and an outer array \cite{pal2010nested}. As shown in Fig.~\ref{fig:arrayArchitectureIllustration}(d), the inner array is a CA with $N_{\rm in}$ antenna pixels, and the outer array is a USA with $N_{\rm ou}$ pixels, where its sparsity level is $\eta = N_{\rm in}+1$, and $N_{\rm in} + N_{\rm ou} = N$ \cite{pal2010nested}. Moreover, the last pixel of the inner array is adjacent to the first pixel of the outer array, with a spacing of $d ={\lambda}/{2}$. In particular, when ${N_{{\rm{in}}}} = 0$, ${N_{{\rm{ou}}}} = 0$, or ${N_{{\rm{ou}}}} = 1$, NA reduces the classic CA. Let the bottom pixel of NA be the reference point, with its position denoted as ${s_{{\rm{ref}}}} = b$. The position of antenna pixel $n$ is
 \begin{equation}
 {s_n} = \left\{ \begin{split}
 &b + \left( {n - 1} \right)d,\ \ \ \ \ \ \ \ \ \ \ \ \ \ \ \ \ \ \ \ \ {\rm if}\ n \le {N_{\rm{in}}},\\
 &b + \left( {n - {N_{\rm{in}}}} \right)\left( {{N_{\rm{in}}} + 1} \right)d - d,\ {\rm otherwise}.
 \end{split} \right.
 \end{equation}
 For any given configuration $\left( {{N_{{\rm{in}}}},{N_{{\rm{o u}}}}} \right)$, the physical dimension is ${D_{{\rm{NA}}}} \left( {N_{\rm in}}  \right) = {N_{{\rm{ou}}}}\left( {{N_{{\rm{in}}}} + 1} \right)d - d$, where $N_{\rm in}$ is used for denoting the antenna configuration, since ${N_{{\rm{ou}}}}$ can be determined based on ${N_{{\rm{ou}}}} = N - {N_{{\rm{in}}}}$.

 The array configuration parameter for NA is the number of inner array's antenna pixels $N_{\rm{in}}$ and reference point position $b$. For any given $\left( {{N_{{\rm{in}}}},{N_{{\rm{ou}}}}} \right)$, the minimum and maximum reference point positions are ${b_{\min }}\left( {{N_{{\rm{in}}}}} \right) =  - \frac{{M - 1}}{2}d$ and ${b_{\max }}\left( {{N_{{\rm{in}}}}} \right) = \frac{{M - 2{N_{{\rm{ou}}}}\left( {{N_{{\rm{in}}}} + 1} \right) + 1}}{2}d$, respectively. Thus, the total realizations for given $\left( {{N_{{\rm{in}}}},{N_{{\rm{ou}}}}} \right)$ are $\left( {{b_{\max }}\left( {{N_{{\rm{in}}}}} \right) - {b_{\min }}\left( {{N_{{\rm{in}}}}} \right)} \right)/d + 1 = M - {N_{{\rm{ou}}}}\left( {{N_{{\rm{in}}}} + 1} \right) + 1$. Besides, to ensure that the physical dimension of the constructed NA is no greater than that of the original array, we have
 \begin{equation}\label{dimensionConstraintNA}
 {N_{{\rm{ou}}}}\left( {{N_{{\rm{in}}}} + 1} \right)d - d \le \left( {M - 1} \right)d.
 \end{equation}

 \begin{proposition} \label{innerArrayElementNumberProposition}
 The feasible region of $N_{\rm in}$ that satisfies \eqref{dimensionConstraintNA} and forms NA is
 \begin{equation}\label{feasibleRegionNin}
 {{\cal S}_{\rm NA}} = \left\{ \begin{split}
 &\left\{ {1,2, \cdots ,N - 2} \right\},\ \ \ \ {\rm if}\ \frac{{{{\left( {N + 1} \right)}^2}}}{4} \le M,\\
 &\left\{ {1, \cdots ,N_{{\rm{in}}}^l} \right\} \cup \left\{ {N_{{\rm{in}}}^u, \cdots ,N - 2} \right\},\\
 &\ \ \ \ \ \ \ \ \ \ \ {\rm if}\ \frac{{{{\left( {N + 1} \right)}^2}}}{4} > M\ {\rm and}\ N_{{\rm{in}}}^l \ge 1,
 \end{split} \right.
 \end{equation}
 where $N_{{\rm{in}}}^l  \triangleq \left\lfloor {\frac{{\left( {N - 1} \right) - \sqrt {{{\left( {N + 1} \right)}^2} - 4M} }}{2}} \right\rfloor$ and $ N_{{\rm{in}}}^u \triangleq  \left\lceil {\frac{{\left( {N - 1} \right) + \sqrt {{{\left( {N + 1} \right)}^2} - 4M} }}{2}} \right\rceil $, respectively.
 \end{proposition}
 \begin{IEEEproof}
 Please refer to Appendix~\ref{proofinnerArrayElementNumberProposition}.
 \end{IEEEproof}

 Thus, the total realizations of NA are ${B_{{\rm{NA}}}} = \sum\nolimits_{{N_{{\rm{in}}}} \in {\cal S}_{\rm NA}}\left( {M - {N_{{\rm{ou}}}}\left( {{N_{{\rm{in}}}} + 1} \right) + 1}\right)$, where the feasible region ${{\cal S}_{\rm NA}}$ has removed the special case of CA. The codebook for NA is
 \begin{equation}\label{codebookNA}
 {{\cal W}_{{\rm{NA}}}} = {\left\{ {{{\bf{w}}_{{\rm{NA}}}}\left( {{N_{{\rm{in}}}},b} \right)} \right\}_{{N_{{\rm{in}}}} \in {{\cal S}_{{\rm{NA}}}},{b_{\min }}\left( {{N_{{\rm{in}}}}} \right) \le b \le {b_{\max }}\left( {{N_{{\rm{in}}}}} \right)}},
 \end{equation}
 where ${{\bf{w}}_{{\rm{NA}}}}\left( {{N_{{\rm{in}}}},b} \right) \in {{\mathbb R}^{N \times 1}} = \left[ {b,b + d, \cdots ,b + {N_{{\rm{in}}}}d,} \right.$ ${\left. {b + 2{N_{{\rm{in}}}}d + d, \cdots ,b + \left( {N - {N_{{\rm{in}}}}} \right)\left( {{N_{{\rm{in}}}} + 1} \right)d - d} \right]^T}$.

\subsubsection{Co-prime Array}
 Another kind of non-uniform sparse array is CPA, which is a superposition of two USAs, with their first antenna pixels overlapped, as illustrated in Fig.~\ref{fig:arrayArchitectureIllustration}(e). The number of pixels of the two USAs are $N_{\rm f}$ and $N_{\rm s}$, respectively, where $N_{\rm f}$ and $N_{\rm s}$ are co-prime integers \cite{vaidyanathan2011sparse}. The adjacent pixels of the first and second USAs are separated by $N_{\rm s}d$ and $N_{\rm f}d$, respectively. Without loss of generality, we assume that $N_{\rm f} < N_{\rm s}$. The total number of antenna pixels for CPA is $N = N_{\rm f} + N_{\rm s}-1$. In particular, when ${N_{\rm{f}}} = 1$ or ${N_{\rm{s}}} = 1$, CPA reduces to the classic CA. Thus, the number of the first and second USA elements should satisfy that ${N_{\rm{f}}} \ge 2$ and ${N_{\rm{s}}} = N + 1 - {N_{\rm{f}}} \ge 2$, yielding ${N_{\rm{f}}} \in \left\{ {2,3, \cdots ,\left\lfloor {\frac{{N + 1}}{2}} \right\rfloor } \right\}$. Let the bottom pixel of CPA be the reference point, with its position denoted as $b$, and ${\bar \Lambda }\left(b\right)$ denote the following set
 \begin{equation}
 \begin{aligned}
 {\bar \Lambda }\left(b\right) = \left\{ {\left. {b + \left( {{n_{\rm{f}}} - 1} \right){N_{\rm{s}}}d} \right|1 \le {n_{\rm{f}}} \le {N_{\rm{f}}}} \right\} \cup \\
 \left\{ {\left. {b + \left( {{n_{\rm{s}}} - 1} \right){N_{\rm{f}}}d} \right|1 \le {n_{\rm{s}}} \le {N_{\rm{s}}}} \right\}.
 \end{aligned}
 \end{equation}
 By re-ordering the elements in ${\bar \Lambda }\left(b\right)$ in an increasing order and denoting the new set as ${\Lambda }\left(b\right)$. The position of antenna pixel $n$ is ${s_n} = {\Lambda _n}\left(b\right)$. For given antenna configuration $\left( {{N_{\rm{f}}},{N_{\rm{s}}}} \right)$, the physical dimension is ${D_{{\rm{CPA}}}} \left( {N_{\rm f}} \right) = \max \left( {\left( {{N_{\rm{f}}} - 1} \right){N_{\rm{s}}}d,\left( {{N_{\rm{s}}} - 1} \right){N_{\rm{f}}}d} \right) = \left( {{N_{\rm{s}}} - 1} \right){N_{\rm{f}}}d$.

 The array configuration parameter for CPA is the number of the first array's antenna pixels $N_{\rm{f}}$ and reference point position $b$. For given $\left( {{N_{\rm{f}}},{N_{\rm{s}}}} \right)$, the minimum and maximum reference point indices are ${b_{\min }}\left( {{N_{\rm{f}}}} \right) = -\frac{M-1}{2}d$ and ${b_{\max }}\left( {{N_{\rm{f}}}} \right) = \frac{{M - 2\left( {{N_{\rm{s}}} - 1} \right){N_{\rm{f}}} - 1}}{2}d$, respectively. Thus, the total realizations for given $\left( {{N_{\rm{f}}},{N_{\rm{s}}}} \right)$ are $\left( {{b_{\max }}\left( {{N_{\rm{f}}}} \right) - {b_{\min }}\left( {{N_{\rm{f}}}} \right)} \right)/d + 1 = M - \left( {{N_{\rm{s}}} - 1} \right){N_{\rm{f}}}$. To ensure that the physical dimension of the constructed CPA is no greater than that of the original array, we should have
 \begin{equation}\label{dimensionConstraintCPA}
 \left( {{N_{\rm{s}}} - 1} \right){N_{\rm{f}}}d \le \left( {M - 1} \right)d.
 \end{equation}

 However, it is difficult to directly obtain the feasible region of ${N_{\rm{f}}}$ satisfying \eqref{dimensionConstraintCPA} and the constraint that ${N_{\rm{f}}}$ and ${N_{\rm{s}}} = N + 1 - {N_{\rm{f}}}$ are co-prime, but it can be obtained by exhaustive search, denoted as ${\cal S}_{\rm CPA}$. Thus, the total realizations of CPA are ${B_{{\rm{CPA}}}} = \sum\nolimits_{{N_{\rm{f}}} \in {\cal S}_{\rm CPA}} {\left( {M - \left( {{N_{\rm{s}}} - 1} \right){N_{\rm{f}}}} \right)}$, and ${\cal S}_{\rm CPA}$ has similarly excluded the special case of CA. The codebook for CPA is
 \begin{equation}\label{codebookUSA}
 {{\cal W}_{{\rm{CPA}}}} = {\left\{ {{{\bf{w}}_{{\rm{CPA}}}}\left( {{N_{\rm{f}}},b} \right)} \right\}_{{N_{\rm{f}}} \in {{\cal S}_{{\rm{CPA}}}},{b_{\min }}\left( {{N_{\rm{f}}}} \right) \le b \le {b_{\max }}\left( {{N_{\rm{f}}}} \right)}},
 \end{equation}
 where ${{\bf{w}}_{\rm{CPA}}}\left( {{N_{\rm{f}}},b} \right) \in {{\mathbb R}^{N \times 1}} =\left[ {{\Lambda _1}\left( b \right),{\Lambda _2}\left( b \right), \cdots ,} \right.$ ${\left. {{\Lambda _N}\left( b \right)} \right]^T}$.

 Finally, by combining the codebooks of the above five array architectures, the overall ACC is given by
 \begin{equation}\label{codebook}
 {\cal W}^{\rm ACC} =  {{{\cal W}_{\rm CA}} \cup {{\cal W}_{\rm USA}} \cup {{\cal W}_{\rm MoA}} \cup {{\cal W}_{\rm NA}} \cup {{\cal W}_{\rm CPA}}},
 \end{equation}
 where $\left| {\cal W}^{\rm ACC} \right| \triangleq {B_{{\rm{CA}}}} + {B_{\rm USA}} + {B_{\rm MoA}} + {B_{\rm NA}} + {B_{\rm CPA}}$ denotes the number of columns of ${\cal W}^{\rm ACC}$, i.e., codebook size. For example, for an XL-ULA with $M = 256$ antenna pixels, when $N = 32$ pixels are activated, the number of realizations for different array architectures are $B_{\rm CA} = 225$, $B_{\rm USA} = 707$, $B_{\rm MoA} = 3504$ ($Z = 8$, $N_m = 4$), $B_{\rm NA} = 1870$, and $B_{\rm CPA} = 655$, respectively. Thus, the total number of codewords is $\left|{{\cal W}^{\rm ACC}}\right| = 6961$.

\section{Array Configuration Training For \\Multi-UE Communication}\label{sectionArrayTrainingCommunication}
 In this section, we study the array configuration training with the designed ACC, by considering the uplink multi-UE communication scenario.{\footnote[1]{The uplink multi-UE communication scenario can be extended to the downlink scenario, by replacing the design of receive beamforming with that of transmit precoding at the BS.}}

 For given array configuration codeword $\bf w$, the channel from UE $k$ to the BS is expressed as
 \begin{equation}\label{ChannelUEkCodeword}
 {{\bf{h}}_k}\left( {\bf{w}} \right) = \sum\limits_{l = 1}^{{L_k}} {{\alpha _{k,l}}{\bf{a}}\left( {{\bf{w}}; {{\bf{q}}_{k,l}} } \right)},
 \end{equation}
 with the array response vector ${\bf{a}}\left( {{\bf{w}}; {{\bf{q}}_{k,l}} } \right)$ given by \cite{lu2024tutorial}
 \begin{equation}\label{arrayResponseVectorkl}
 \begin{aligned}
 {\bf{a}}\left( {{\bf{w}};{{\bf{q}}_{k,l}}} \right) &= \left[ {\frac{{{r_{k,l}}}}{{{r_{k,l,1}}}}{e^{ - {\rm{j}}\frac{{2\pi }}{\lambda }\left( {{r_{k,l,1}} - {r_{k,l}}} \right)}}, \cdots ,} \right.\\
 &\ \ \ \ \ \ \ \ \ \ \ {\left. {\frac{{{r_{k,l}}}}{{{r_{k,l,N}}}}{e^{ - {\rm{j}}\frac{{2\pi }}{\lambda }\left( {{r_{k,l,N}} - {r_{k,l}}} \right)}}} \right]^T},
 \end{aligned}
 \end{equation}
 where ${r_{k,l}}$ denotes the link distance between ${{\bf{q}}_{k,l}}$ and the origin, and ${r_{k,l,n}} = \left\| {{{\bf{q}}_{k,l}} - {{\bf{w}}_n}} \right\|$ denotes the distance between ${{\bf{q}}_{k,l}}$ and ${{{\bf{w}}_n}}$, with ${{{\bf{w}}_n}}$ being the position of antenna pixel $n$.

 Denote by $e_i$ and ${{P_i}}$ the information-bearing symbol and transmit power of UE $i$, respectively, $1 \le i \le K$. The received signal at the BS is
 \begin{equation}\label{receivedSignalBS}
 {\bf{y}}\left( {\bf{w}} \right) = \sqrt {{P_k}} {{\bf{h}}_k}\left( {\bf{w}} \right){e_k} + \sum\limits_{i = 1,i \ne k}^K {\sqrt {{P_i}} {{\bf{h}}_i}\left( {\bf{w}} \right){e_i}}  + {\bf{n}},
 \end{equation}
 where ${\bf{n}} \sim {\cal CN}\left( {{\bf{0}},{\sigma ^2}{{\bf{I}}_N}} \right)$ denotes the additive white Gaussian noise (AWGN) with zero mean and covariance matrix ${\sigma ^2}{{\bf{I}}_N}$.

 Moreover, the receive beamforming ${{\bf{v}}_k}$ is applied to detect the signal of UE $k$, where $\left\| {{{\bf{v}}_k}} \right\| = 1$. The resulting SINR for UE $k$ is
 \begin{equation}\label{receivedSINRUEkCA}
 \begin{aligned}
 {\gamma _k}\left( {{\bf{w}},{{\bf{v}}_k}} \right) &= \frac{{{{\bar P}_k}{{\left| {{\bf{v}}_k^H{{\bf{h}}_k}\left( {\bf{w}} \right)} \right|}^2}}}{{\sum\limits_{i = 1,i \ne k}^K {{{\bar P}_i}{{\left| {{\bf{v}}_k^H{{\bf{h}}_i}\left( {\bf{w}} \right)} \right|}^2}}  + 1}}\\
 &= \frac{{{{\bar P}_k}{\bf{v}}_k^H{{\bf{h}}_k}\left( {\bf{w}} \right){\bf{h}}_k^H\left( {\bf{w}} \right){{\bf{v}}_k}}}{{{\bf{v}}_k^H{{\bf{C}}_k}\left( {\bf{w}} \right){{\bf{v}}_k}}},
 \end{aligned}
 \end{equation}
 where ${{\bar P}_i} \triangleq {P_i}/{\sigma ^2}$, and ${{\bf{C}}_k}\left( {\bf w} \right) \in {\mathbb C}^{N \times N} \triangleq {\bf{I}}_N + \sum\nolimits_{i = 1,i \ne k}^K {{{\bar P}_i}{{\bf{h}}_i}\left( {\bf w} \right){\bf{h}}_i^H\left( {\bf w} \right)} $ is the interference-plus-noise covariance matrix of UE $k$. Since ${\gamma _k}\left( {{\bf{w}},{{\bf{v}}_k}} \right)$ constitutes a generalized Rayleigh quotient w.r.t. ${{{\bf{v}}_k}}$, the optimal ${{{\bf{v}}_k}}$ to maximize \eqref{receivedSINRUEkCA} is
 ${{\bf{v}}_k^{\star}} = {\bf{C}}_k^{ - 1}\left( {\bf w} \right){{\bf{h}}_k}\left( {\bf w} \right)/\left\| {{\bf{C}}_k^{ - 1}\left( {\bf w} \right){{\bf{h}}_k}\left( {\bf w} \right)} \right\|$ \cite{molisch2012wireless}. The resulting SINR for UE $k$ is
 \begin{equation}\label{receivedSINRUEk}
 {\gamma _k}\left( {\bf w} \right) = {{\bar P}_k}{\bf{h}}_k^H\left( {\bf w} \right){\bf{C}}_k^{ - 1}\left( {\bf w} \right){{\bf{h}}_k}\left( {\bf w} \right),
 \end{equation}
 where the dimension of matrix inversion is $N \times N$.

 Let ${{\bf{H}}_k}\left( {\bf w} \right) \in {{\mathbb C}^{N \times \left( {K - 1} \right)}} \triangleq \left[ {{{\bf{h}}_1}\left( {\bf w} \right), \cdots ,{{\bf{h}}_{k - 1}}\left( {\bf w} \right),} \right.$ $\left. {{{\bf{h}}_{k + 1}}\left( {\bf w} \right), \cdots ,{{\bf{h}}_K}\left( {\bf w} \right)} \right]$, and ${{\bf{P}}_k} \in {{\mathbb R}^{\left( {K - 1} \right) \times \left( {K - 1} \right)}} \triangleq {\rm{diag}}\left( {{{\bar P}_1}, \cdots ,{{\bar P}_{k - 1}},{{\bar P}_{k + 1}}, \cdots ,{{\bar P}_K}} \right)$. By invoking the Woodbury matrix identity ${\left( {{\bf{A}} + {\bf{BCD}}} \right)^{ - 1}} = {{\bf{A}}^{ - 1}} - {{\bf{A}}^{ - 1}}{\bf{B}}{\left( {{{\bf{C}}^{ - 1}} + {\bf{D}}{{\bf{A}}^{ - 1}}{\bf{B}}} \right)^{ - 1}}{\bf{D}}{{\bf{A}}^{ - 1}}$, ${\bf{C}}_k^{ - 1}\left( {\bf w} \right)$ can be expressed as
 \begin{equation}\label{covarianceMatrix}
 \begin{aligned}
 &{\bf{C}}_k^{ - 1}\left( {\bf w} \right) = {\left( { {\bf{I}}_N + {{\bf{H}}_k}\left({\bf w} \right){{\bf{P}}_k}{\bf{H}}_k^H\left( {\bf w} \right)} \right)^{ - 1}}\\
 &\ \ = {\bf{I}}_N - {{\bf{H}}_k}\left( {\bf w} \right){\left( {{\bf{P}}_k^{ - 1} + {\bf{H}}_k^H\left( {\bf w} \right){{\bf{H}}_k}\left( {\bf w} \right)} \right)^{ - 1}}{\bf{H}}_k^H\left( {\bf w} \right).
 \end{aligned}
 \end{equation}
 Then, the resulting SINR for UE $k$ can be equivalently expressed as
 \begin{equation}\label{alternativeReceivedSINRUEk}
 \begin{aligned}
 &{\gamma _k}\left( {\bf{w}} \right) = {{\bar P}_k}\left( {{{\left\| {{{\bf{h}}_k}\left( {\bf{w}} \right)} \right\|}^2} - {\bf{h}}_k^H\left( {\bf{w}} \right){{\bf{H}}_k}\left( {\bf{w}} \right)} \right.\\
 &\ \ \  \left. { \times {{\left( {{\bf{P}}_k^{ - 1} + {\bf{H}}_k^H\left( {\bf{w}} \right){{\bf{H}}_k}\left( {\bf{w}} \right)} \right)}^{ - 1}}{\bf{H}}_k^H\left( {\bf{w}} \right){{\bf{h}}_k}\left( {\bf{w}} \right)} \right).
 \end{aligned}
 \end{equation}
 It is observed that the matrix inversion dimension is $\left( {K - 1} \right) \times \left( {K - 1} \right)$. Thus, depending on the relationship between $N$ and $K-1$, the SINR expressions in \eqref{receivedSINRUEk} or \eqref{alternativeReceivedSINRUEk} can be used for reducing the computational complexity.

 The achievable sum rate of all the $K$ UEs is
 \begin{equation}\label{sumRateCA}
 R\left( {\bf{w}} \right) = \sum\limits_{k = 1}^K {{{\log }_2}\left( {1 + {\gamma _k}\left( {\bf{w}} \right)} \right)}.
 \end{equation}



 For the considered multi-UE communication, the sum rate is adopted as the utility function, and array configuration training aims to maximize the sum rate. The optimization problem \eqref{optimizationProblemACC} becomes
 \begin{equation}\label{arrayTrainingCommunicationProblem}
 \begin{aligned}
 \mathop {\max }\limits_{{\bf w}  }&\ \  R\left( {\bf{w}} \right)\\
 {\rm{s.t.}}&\ \  {\bf w} \in {\cal W}^{\rm ACC}.
 \end{aligned}
 \end{equation}

 \subsection{Exhaustive Scanning Over ACC}
 Given the designed ACC in Section~\ref{sectionACCDesign}, one straightforward scheme to find the optimal array configuration codeword is the exhaustive scanning. Specifically, by comparing the sum rate of all the $\left|{{\cal W}^{\rm ACC}}\right|$ codewords, the optimal codeword corresponds to the one with the maximum sum rate. The overhead of exhaustive scanning is ${\cal O} \left(\left|{{\cal W}^{\rm ACC}}\right| \right)$. For instance, when $M = 256$ and $N = 32$, the training overhead is 6961, given below equation \eqref{codebook}. By contrast, for exhaustive search in AS, the number of antenna pixel position combinations is $\binom{256}{32} \approx 5.824 \times {10^{40}}$, i.e., the training overhead is $5.824 \times {10^{40}}$, which is prohibitive in practical scenarios. 

 \subsection{Two-Stage Scanning Over ACC}\label{subsectionTwoStageScanning}
 In this subsection, to further reduce the training overhead of exhaustive scanning scheme, we propose an efficient two-stage array configuration training scheme, i.e., the array- and pixel-level scanning.
 \begin{figure}[!t]
 \centering
 \centerline{\includegraphics[width=3.0in,height=2.5in]{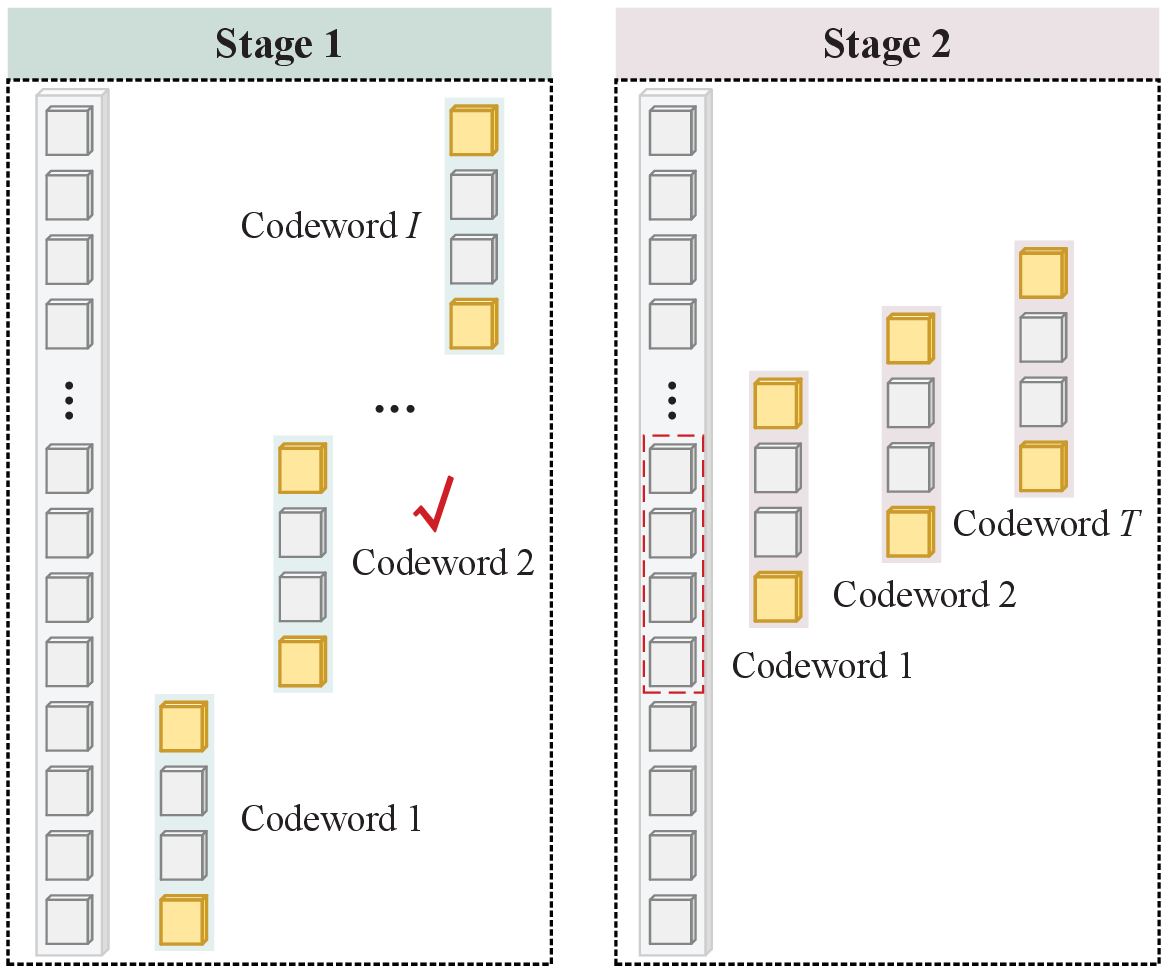}}
 \caption{An illustration of the two-stage array configuration training.}
 \label{fig:twoStageSearchIllustration}
 \end{figure}
 \subsubsection{Array-Level Scanning in Stage 1}
 In stage 1, the coarse array-level scanning is performed. Specifically, the number of consecutive antenna pixels required to form any given array architecture is ${{\bar N}_{a}} = {D_a}/d + 1$, where ${D_a}$ denotes the physical dimension of array architecture $a$, with $a \in \left\{ {{\rm CA},{\rm USA},{\rm MoA},{\rm NA},{\rm CPA}} \right\}$. Thus, for the XL-ULA with $M$ elements, the total number of this array architectures can be constructed without intersection is ${I_a} = \left\lfloor {M/{{\bar N}_{a}}} \right\rfloor $, with each realization corresponding to one codeword to be scanned in stage 1, as illustrated in Fig.~\ref{fig:twoStageSearchIllustration}. This thus yields the following reduced codebooks for different array architectures, which are summarized in Table \ref{table:codebookStage1}.

 \begin{table*}[!t]
	\centering
	\caption{Reduced Codebooks for Different Array Architectures in Stage 1}\label{table:codebookStage1}
   {
		\begin{tabular}{|m{1.65cm}<{\centering}|m{10.8cm}<{\centering}|m{4.5cm}<{\centering}| }
			\hline
			{\bf Array Architectures} &
			{\bf Codebook in Stage 1 } &
            {\bf Codebook Size }
 \\ \hline\hline
			{\bf CA}                         & {\begin{tabular}[c]{@{}c@{}}${\cal W}_{\rm CA}^{\left( 1 \right)} = {\left\{ {{{\bf{w}}_{{\rm{CA}}}}\left( b \right)} \right\}_{b \in {\cal B}_{{\rm{CA}}}^{\left( 1 \right)}}}$\\ ${{\cal B}_{{\rm{CA}}}^{\left( 1 \right)}} = \left\{ {{b_{\min }},{b_{\min }} + {{\bar N}_{\rm CA}}d, \cdots ,{b_{\min }} + }  {\left( {{I_{{\rm{CA}}}} - 1} \right){{\bar N}_{{\rm{CA}}}}d} \right\}$\end{tabular}}  &   $\left| {{\cal W}_{\rm CA}^{\left( 1 \right)}} \right| = {I_{\rm CA}}$
\\ \hline
{\bf USA}                         &  {\begin{tabular}[c]{@{}c@{}}${{\cal W}_{{\rm{USA}}}^{\left( 1 \right)}} = {\left\{ {{{\bf{w}}_{{\rm{USA}}}}\left( {\eta ,b} \right)} \right\}_{2 \le \eta  \le {\eta _{\max }},b \in {\cal B}_{{\rm{USA}}}^{\left( 1 \right)}\left( \eta  \right) } }$,\\ ${\cal B}_{{\rm{USA}}}^{\left( 1 \right)}\left( \eta  \right) =\left\{ {{b_{\min }},{b_{\min }} + {{\bar N}_{{\rm{USA}}}}\left( \eta  \right)d, \cdots ,{b_{\min }} + }  {\left( {{I_{{\rm{USA}}}}\left( \eta  \right) - 1} \right){{\bar N}_{{\rm{USA}}}}\left( \eta  \right)d} \right\}$\end{tabular}}    &    $\left| {{\cal W}_{{\rm{USA}}}^{\left( 1 \right)}}\right| = \sum\nolimits_{\eta  = 2}^{{\eta _{\max }}} {{I_{{\rm{USA}}}}\left( \eta  \right)} $
\\ \hline
{\bf MoA}                         & {\begin{tabular}[c]{@{}c@{}}$ {\cal W}_{{\rm{MoA}}}^{\left( 1 \right)} = {\left\{ {{{\bf{w}}_{{\rm{MoA}}}}\left( {\Gamma ,b} \right)} \right\}_{{N_m} + 1 \le \Gamma  \le {\Gamma _{\max }},b \in {\cal B}_{\rm MoA}^{\left( 1 \right)}\left( \Gamma  \right)}}$,\\ ${\cal B}_{\rm MoA}^{\left( 1 \right)}\left( \Gamma  \right) = \left\{ {{b_{\min }},{b_{\min }} + {{\bar N}_{{\rm{MoA}}}}\left( \Gamma  \right)d, \cdots ,{b_{\min }} + }  {\left( {{I_{{\rm{MoA}}}}\left( \Gamma  \right) - 1} \right){{\bar N}_{{\rm{MoA}}}}\left( \Gamma  \right)d} \right\}$\end{tabular}}   &    $\left| {{\cal W}_{{\rm{MoA}}}^{\left( 1 \right)}} \right| = \sum\nolimits_{\Gamma  = {N_m} + 1}^{{\Gamma _{\max }}} {{I_{{\rm{MoA}}}}\left( \Gamma  \right)} $
\\ \hline
{\bf NA}                         & {\begin{tabular}[c]{@{}c@{}}${\cal W}_{{\rm{NA}}}^{\left( 1 \right)} = {\left\{ {{{\bf{w}}_{{\rm{NA}}}}\left( {{N_{{\rm{in}}}},b} \right)} \right\}_{{N_{{\rm{in}}}} \in {{\cal S}_{{\rm{NA}}}},b \in {\cal B}_{{\rm{NA}}}^{\left( 1 \right)}\left( {{N_{{\rm{in}}}}} \right)}}$,\\ ${\cal B}_{{\rm{NA}}}^{\left( 1 \right)}\left( {{N_{{\rm{in}}}}} \right)= \left\{ {{b_{\min }},{b_{\min }} + {{\bar N}_{{\rm{NA}}}}\left( {{N_{\rm in}}} \right)d, \cdots ,{b_{\min }} + }  {\left( {{I_{{\rm{NA}}}}\left( {{N_{{\rm{in}}}}} \right) - 1} \right){{\bar N}_{{\rm{NA}}}}\left( {{N_{\rm in}}} \right)d} \right\}$\end{tabular}}  &   $\left| {{\cal W}_{{\rm{NA}}}^{\left( 1 \right)}} \right| = \sum\nolimits_{{N_{{\rm{in}}}} \in {{\cal S}_{\rm{NA}}}} {{I_{{\rm{NA}}}}\left( {{N_{{\rm{in}}}}} \right)} $
\\ \hline
{\bf CPA}                         & {\begin{tabular}[c]{@{}c@{}}$ {\cal W}_{{\rm{CPA}}}^{\left( 1 \right)} = {\left\{ {{{\bf{w}}_{{\rm{CPA}}}}\left( {{N_{\rm{f}}},b} \right)} \right\}_{{N_{\rm{f}}} \in {{\cal S}_{{\rm{CPA}}}},b \in {\cal B}_{\rm CPA}^{\left( 1 \right)}\left( {{N_{\rm{f}}}} \right)}}$,\\ ${\cal B}_{\rm CPA}^{\left( 1 \right)}\left( {{N_{\rm{f}}}} \right) = \left\{ {{b_{\min }},{b_{\min }} + {{\bar N}_{{\rm{CPA}}}}\left( {{N_{\rm{f}}}} \right)d, \cdots ,{b_{\min }} + }  {\left( {{I_{{\rm{CPA}}}}\left( {{N_{\rm{f}}}} \right) - 1} \right){{\bar N}_{{\rm{CPA}}}}\left( {{N_{\rm{f}}}} \right)d} \right\}$\end{tabular}}  & $\left| {{\cal W}_{{\rm{CPA}}}^{\left( 1 \right)}} \right| = \sum\nolimits_{{N_{\rm{f}}} \in {{\cal S}_{{\rm{CPA}}}}} {{I_{{\rm{CPA}}}}\left( {{N_{\rm{f}}}} \right)} $
\\ \hline
		\end{tabular}
	}
\end{table*}

 Then, the BS performs the array-level scanning to find the best codeword for each array architecture, given by
 \begin{equation}\label{bestArrayCodewordStage1}
 {\bf{w}}_a^{\left(1\right)} = \arg \mathop {\max }\limits_{{\bf{w}} \in {\cal W}_a^{\left( 1 \right)}} R\left( {\bf{w}} \right).
 \end{equation}

 \subsubsection{Pixel-Level Scanning in Stage 2}
 With the obtained array-level codeword, a finer scanning over the reference points is then implemented. It is worth mentioning that the codeword obtained in \eqref{bestArrayCodewordStage1} may incorporate the information of reference point position and architecture-specific parameter (i.e., $\eta$, $\Gamma$, $N_{\rm in}$, and $N_{\rm f}$ for USA, MoA, NA, and CPA, respectively). To this end, let ${\varphi _a}$ denote the architecture-specific parameter for array architecture $a$, given by
 \begin{equation}
 {\varphi _a} = \left\{ \begin{aligned}
 &\emptyset, \ a = {\rm{CA}},\\
 &\eta ,\ a = {\rm{USA}},\\
 &\Gamma ,\ a = {\rm{MoA}},\\
 &{N_{{\rm{in}}}},\ a = {\rm{NA}},\\
 &{N_{\rm{f}}},\ a = {\rm{CPA}}.
 \end{aligned} \right.
 \end{equation}

 Note that since there is no architecture-specific parameter required for the classic CA, the notation $\emptyset$ is used. Let $( { \varphi _a^{\left( 1 \right)}},{b_a^{\left( 1 \right)}} )$ denote the pair of the architecture-specific parameter and reference point position corresponding to ${\bf{w}}_a^{\left(1\right)}$ for array architecture $a$. Then, in stage 2, the reference point is scanning pixel-by-pixel within the array formed by $( { \varphi _a^{\left( 1 \right)}},{b_a^{\left( 1 \right)}} )$. For example, as illustrated in Fig.~\ref{fig:twoStageSearchIllustration}, when codeword 2 is determined in stage 1, a successive codeword scanning is performed by changing the reference point position as any pixel position within the array occupied by codebook 2 in stage 1, except for the first one.

 Let ${T_a}( {\varphi _a^{\left( 1 \right)}})$ denote the number of candidate codewords in stage 2, given by
 \begin{equation}
 \begin{aligned}
 &T_a( {\varphi _a^{\left( 1 \right)}}) = \min \left( {{{\bar N}_a}( {\varphi _a^{\left( 1 \right)}} ) - 1,} \right.\\
 &\ \ \ \ \ \ \ \ \ \ \ \ \ \ \left. {\frac{{M - 1}}{2} - \left( {{{\bar N}_a}( {\varphi _a^{\left( 1 \right)}}) - 1} \right) - b_a^{\left( 1 \right)}/d} \right),
 \end{aligned}
 \end{equation}
 where the second term is due to the fact that the ${{\bar N}_a}( {\varphi _a^{\left( 1 \right)}} )$-th antenna pixel cannot exceed the last pixel of the XL-ULA. In particular, for the CA, ${{\bar N}_a}( {\varphi _a^{\left( 1 \right)}} ) = {{\bar N}_{{\rm{CA}}}}$. Then, the codebook to be scanned for array architecture $a$ in stage 2 is
 \begin{equation}\label{codebookStage2}
 {\cal W}_a^{\left( 2 \right)} = {\left\{ {{{\bf{w}}_a}\left( {\varphi _a^{\left( 1 \right)},b} \right)} \right\}_{b \in {\cal B}_a^{\left( 2 \right)}}},
 \end{equation}
 with
 \begin{equation}\label{referencePointStage2}
 {\cal B}_a^{\left( 2 \right)} = \left\{ {b_a^{\left( 1 \right)} + d,b_a^{\left( 1 \right)} + 2d, \cdots ,b_a^{\left( 1 \right)} + {T_a}( {\varphi _a^{\left( 1 \right)}})d} \right\}.
 \end{equation}

 With \eqref{codebookStage2} and \eqref{referencePointStage2}, the pixel-level scanning is performed to obtain the best codeword separately within each array architecture, given by
 \begin{equation}\label{bestArrayCodewordStage2}
 {\bf{w}}_a^{\left( 2 \right)} = \arg \mathop {\max }\limits_{{\bf{w}} \in {\cal W}_a^{\left( 2 \right)}} R\left( {\bf{w}} \right).
 \end{equation}

 Then, by comparing the sum rates under the codewords obtained in \eqref{bestArrayCodewordStage2}, the best codeword within the overall ACC can be determined, given by 
 \begin{equation}\label{bestCodewordStage2}
 {{\bf{w}}^ \star } = \arg \mathop {\max }\limits_{{\bf{w}} \in \left\{ {{\bf{w}}_{{\rm{CA}}}^{(2)},{\bf{w}}_{{\rm{USA}}}^{(2)},{\bf{w}}_{{\rm{MoA}}}^{(2)},{\bf{w}}_{{\rm{NA}}}^{(2)},{\bf{w}}_{{\rm{CPA}}}^{(2)}} \right\}} R\left( {\bf{w}} \right).
 \end{equation}

 \begin{algorithm}[t]
 \caption{Two-Stage Array Configuration Training Scheme Over ACC}
 \label{alg1}
 \begin{algorithmic}[1]
 \STATE \textbf{Input:} The ACC and corresponding array physical dimension $D_a$, $a \in \left\{ {{\rm CA},{\rm USA},{\rm MoA},{\rm NA},{\rm CPA}} \right\}$.
 \STATE Calculate ${{\bar N}_{a}} = {D_a}/d + 1$ and ${I_a} = \left\lfloor {M/{{\bar N}_{a}}} \right\rfloor $.
 \STATE Obtain the codebooks for different array architectures in stage 1 based on Table \ref{table:codebookStage1}, and find the best codeword for each array architecture in stage 1 based on \eqref{bestArrayCodewordStage1}.
 \STATE Obtain the codebooks for different array architectures in stage 2 based on \eqref{codebookStage2} and \eqref{referencePointStage2}, and find the best codeword for each array architecture in stage 2 based on \eqref{bestArrayCodewordStage2}.
 \STATE Obtain the best codeword over ACC ${\bf w}^{\star}$ based on \eqref{bestCodewordStage2}.
 \STATE \textbf{Output:} The best codeword ${\bf w}^{\star}$.
 \end{algorithmic}
 \end{algorithm}

 The main procedures of the proposed two-stage array configuration training scheme over ACC are summarized in Algorithm~\ref{alg1}. Moreover, the overhead of the two-stage training scheme is ${\cal O}( {\sum\nolimits_a {| {{\cal W}_a^{\left( 1 \right)}} | +  | {{\cal W}_a^{\left( 2 \right)}} |} })$. Note that the architecture-specific parameters may vary for different channels, the worst-case training overhead is considered. In particular, when ${N_a} = M/2$, the codebooks to be scanned in stage 2 is given by $M/2 -1$ for all array architectures. Thus, the worst-case training overhead in stage 2 for array architecture $a$ is ${{\cal O}_{{\rm{worst}}}}( {| {{\cal W}_a^{\left( 2 \right)}}|}) = M/2 -1$. By considering the same example where $N = 32$ antenna pixels are activated from $M = 256$ pixels, we have $| {{\cal W}_{\rm CA}^{\left( 1 \right)}} | = 8$, $| {{\cal W}_{\rm USA}^{\left( 1 \right)}} | = 12$, $| {{\cal W}_{\rm MoA}^{\left( 1 \right)}} | = 185$, $| {{\cal W}_{\rm NA}^{\left( 1 \right)}} | = 148$, $| {{\cal W}_{\rm CPA}^{\left( 1 \right)}} | = 132$. Thus, the worst-case training overhead of the proposed two-stage scanning scheme is $8 + 12 + 185 + 148 + 132 + 5 \times (M/2 -1) = 1120$, which significantly reduces the overhead compared to exhaustive scanning. It is also worth mentioning that the actual overhead ${\cal O}( {\sum\nolimits_a {| {{\cal W}_a^{\left( 1 \right)}} | +  | {{\cal W}_a^{\left( 2 \right)}} |} })$ is in general smaller than the worst-case value. Moreover, the two-stage array configuration training scheme can be performed in a parallel manner to accelerate the training speed, say, parallel training across different array architectures.

\subsection{Benchmark Scheme: Greedy AS}
 In this subsection, we propose a benchmark scheme of greedy AS for the multi-UE communication.

 Let ${\bf{u}} \in {{\mathbb R}^{M \times 1}}$ denote the AS indicator vector, where ${u_m} \in \left\{ {1,0} \right\}$, $1 \le m\le M$, and ${u_m} = 1$ (${u_m} = 0$) indicates that antenna pixel $m$ is selected (not selected). Besides, let $\Omega  \triangleq \left\{ {m:{u_m} = 1} \right\}$ denote the set consists of selected antenna pixels. Then, the AS matrix ${\bf{U}} \in {{\mathbb R}^{N \times M}}$ can be correspondingly constructed by setting ${\left[ {\bf{U}} \right]_{n,{\Omega _n}}} = 1$ and other elements of the $n$-th row to zero. For example, for an antenna array with $M=4$ elements, if the second and third elements are selected, we have ${\bf{u}} = {\left[ {0,1,1,0} \right]^T}$, $\Omega  = \left\{ {2,3} \right\}$, and
 \begin{equation}
 {\bf{U}} = \left[ \begin{aligned}
 0,1,0,0\\
 0,0,1,0
 \end{aligned} \right].
 \end{equation}

 With the selection matrix $\bf{U}$, the received signal at the BS becomes
 \begin{equation}\label{receivedSignalBSAfterAS}
 {\bf{\tilde y}} = {\bf{Uy}} = \sqrt {{P_k}} {{{\bf{\tilde h}}}_k}{e_k} + \sum\limits_{i = 1,i \ne k}^K {\sqrt {{P_i}} {{{\bf{\tilde h}}}_i}{e_i}}  + {\bf{\tilde n}},
 \end{equation}
 where ${{{\bf{\tilde h}}}_k} \in {{\mathbb C}^{N \times 1}} = {\bf{U}}{{\bf{h}}_k}$ denotes the truncated channel of UE $k$, $1 \le k \le K$, and ${\bf{\tilde n}} = {\bf{U n}}$. Let ${{\bf{H}}_k} \in {{\mathbb C}^{N \times \left( {K - 1} \right)}} \triangleq [ {{{{\bf{\tilde h}}}_1}, \cdots ,{{{\bf{\tilde h}}}_{k - 1}},{{{\bf{\tilde h}}}_{k + 1}}, \cdots ,{{{\bf{\tilde h}}}_K}} ]$. Similar to \eqref{alternativeReceivedSINRUEk}, the resulting SINR of UE $k$ after applying the optimal receive beamforming ${\bf{v}}_k^ \star  = {\bf{C}}_k^{ - 1}{{{\bf{\tilde h}}}_k}/\left\| {{\bf{C}}_k^{ - 1}{{{\bf{\tilde h}}}_k}} \right\|$, with $ {{\bf{C}}_k} \in {{\mathbb C}^{N \times N}} \triangleq {\bf{I}}_N + \sum\nolimits_{i = 1,i \ne k}^K {{{\bar P}_i}{{{\bf{\tilde h}}}_i}{\bf{\tilde h}}_i^H} $, is
  \begin{equation}\label{alternativeReceivedSINRUEkAS}
 {\gamma _k} = {{\bar P}_k}\left( {{{\left\| {{{{\bf{\tilde h}}}_k}} \right\|}^2} - {\bf{\tilde h}}_k^H{{\bf{H}}_k}{{\left( {{\bf{P}}_k^{ - 1} + {\bf{H}}_k^H{{\bf{H}}_k}} \right)}^{ - 1}}{\bf{H}}_k^H{{{\bf{\tilde h}}}_k}} \right).
 \end{equation}

 The AS vector ${\bf{u}}$ is then optimized to maximize the sum rate. The optimization problem can be formulated as
 \begin{align}\label{OptimizationProblemAS2}
 \left( {\rm{P{\text -}AS}} \right)\ \mathop {\max }\limits_{\bf{u}} &\ \ \sum\limits_{k = 1}^K {{{\log }_2}\left( {1 + {\gamma _k}} \right)}\\
 {\rm{s.t.}}&\ {u_{m}} \in \left\{ {0,1} \right\},\ \forall m, \tag{\ref{OptimizationProblemAS2}a}\\
 &\ {\left\| {\bf{u}} \right\|_0} = N.\tag{\ref{OptimizationProblemAS2}b}
 \end{align}

 To solve the non-convex optimization problem (P-AS), we propose a greedy search scheme, where in each step, one antenna pixel is selected to maximize the achievable sum rate. Specifically, let $\gamma _k^{\left( {n - 1} \right)}$ represent the resulting SINR of UE $k$ at the $\left(n-1\right)$-th step, and $\gamma _{k,m}^{\left( n \right)}$ represent its SINR after selecting $m$ at the $n$-th step. Let ${\bf{\tilde h}}_k^{\left( {n - 1} \right)}$ represent the channel of UE $k$ after the $\left( n-1\right)$-th step. When antenna pixel $m$ is selected at the $n$-th step, it follows that ${\bf{\tilde h}}_k^{\left( n \right)} = {[ {{{( {{\bf{\tilde h}}_k^{\left( {n - 1} \right)}} )^T}},{h_{k,m}}} ]^T}$, where $h_{k,m}$ represents the $m$-th element of ${\bf h}_k$. Besides, let ${{\bf{d}}_{k,m}} \in {{\mathbb C}^{\left( {K - 1} \right) \times 1}} \triangleq {\left[ {{h_{1,m}}, \cdots ,{h_{k - 1,m}},{h_{k + 1,m}}, \cdots ,{h_{K,m}}} \right]^H}$ and ${{\bf{G}}_k^{\left( n \right)}} \triangleq {( {{\bf{P}}_k^{ - 1} + {( {{\bf{H}}_k^{\left( n \right)}} )^H}{\bf{H}}_k^{\left( n \right)}} )^{ - 1}}$. In order to leverage the result of the previous step, we further express the SINR of each UE in an incremental manner, as shown in the following proposition.
 \begin{proposition}\label{incrementalExpressionofSINRProposition}
 The SINR of UE $k$ after selecting antenna pixel $m$ at the $n$-th step can be expressed as
 \begin{equation}\label{incrementalExpressionofSINRUEk}
 \gamma _{k,m}^{\left( n \right)} =   \gamma _k^{\left( {n - 1} \right)} + {{\bar P}_k}\frac{{{{\left| {h_{k,m}^ *  - {{\left( {{\bf{\tilde h}}_k^{\left( {n - 1} \right)}} \right)}^H}{\bf{H}}_k^{\left( {n - 1} \right)}{\bf{G}}_k^{\left( {n - 1} \right)}{{\bf{d}}_{k,m}}} \right|}^2}}}{{1 + {\bf{d}}_{k,m}^H{\bf{G}}_k^{\left( {n - 1} \right)}{{\bf{d}}_{k,m}}}}.
 \end{equation}

 \end{proposition}

 \begin{IEEEproof}
 Please refer to Appendix~\ref{proofIncrementalExpressionofSINRProposition}.
 \end{IEEEproof}

 Proposition \ref{incrementalExpressionofSINRProposition} shows that $\gamma _{k,m}^{\left( n \right)}$ can be expressed as the SINR at the $\left(n-1\right)$-th step plus an increment due to the selection of antenna pixel $m$. Such an expression reuses the previous result and avoids the matrix inversion in \eqref{alternativeReceivedSINRUEk}, thus reducing the computational complexity. Moreover, with \eqref{incrementalExpressionofSINRUEk}, the sum rate $R_m^{\left( n \right)}$ is given by $R_m^{\left( n \right)} = \sum\nolimits_{k = 1}^K {{{\log }_2}( {1 + \gamma _{k,m}^{\left( n \right)}} )}$.
 The best antenna pixel is selected at the $n$-th step such that
 \begin{equation}
 {m^{\left( n \right)}} = \arg \mathop {\max }\limits_m R_m^{\left( n \right)}.
 \end{equation}

 We summarize the main procedures of the proposed greedy AS scheme in Algorithm~\ref{alg2}. Note that the number of candidate antenna pixels at the $n$-th step is $M - n +1$, and the total number of computations is $\sum\nolimits_{n = 1}^N {\left( {M - n + 1} \right)}  = \left( {2M - N + 1} \right)N/2$. For example, when $M = 256$ and $N = 32$, we have $\left( {2M - N + 1} \right)N/2 = 7696$. Moreover, the complexity for calculating \eqref{incrementalExpressionofSINRUEk} at the $n$-th step is ${\cal O}( {\left( {M - n +1 } \right)K\max \{ {\left( {n - 1} \right)\left( {K - 1} \right),{{\left( {K - 1} \right)}^2}} \}} )$. As a result, the overall complexity of Algorithm~\ref{alg2} is ${\cal O}( {\sum\nolimits_{n = 1}^N {\left( {M - n +1} \right)K\max \{ {\left( {n - 1} \right)\left( {K - 1} \right),{{\left( {K - 1} \right)}^2}} \}} } )$.

 \begin{algorithm}[t]
 \caption{Proposed Greedy AS Scheme}
 \label{alg2}
 \begin{algorithmic}[1]
 \STATE Initialize ${\cal M} = \left\{ {1, \cdots ,M} \right\}$, $\Omega  = \emptyset$, $\gamma _k^{\left( 0 \right)} = 0$, and ${\bf{G}}_k^{\left( 0 \right)} = {{\bf P}_k}$, $\forall k$.
 \STATE \textbf{for} $n = 1:N$ \textbf{do}
 \STATE \quad Calculate $\gamma _{k,m}^{\left( n \right)}$ based on \eqref{incrementalExpressionofSINRUEk}, $\forall k$, $m \in {\cal M}$.
 \STATE \quad Obtain the sum rate $R_m^{\left( n \right)}$, $\forall m \in {\cal M}$.
 \STATE \quad ${m^{\left(n\right)} } = \arg \mathop {\max }\limits_{m \in {\cal M}} R_m^{\left( n \right)}$.
 \STATE \quad ${\Omega} = {\Omega} \cup  {{m^ {\left(n\right)} }} $.
 \STATE \quad ${\cal M} = {\cal M}\backslash {m^ {\left(n\right)} }$.
 \STATE \textbf{end for}
 \STATE \textbf{Output:} The set of selected antenna pixels $\Omega$.
 \end{algorithmic}
 \end{algorithm}

\section{Array Configuration Training For \\Wireless Localization}\label{sectionArrayTrainingLocalization}
 In this section, we study the array configuration training for the wireless localization scenario, where the number of localization UEs is $K$.

 For given codeword $\bf w$, the LoS channel from localization UE $k$ to the BS is
 \begin{equation}\label{localizationChannel}
 {{\bf{h}}_k}\left( {\bf{w}} \right) = {\alpha _k}{\bf{a}}\left( {{\bf{w}};{{\bf{q}}_k}} \right),
 \end{equation}
 where ${\bf{a}}\left( {{\bf{w}};{{\bf{q}}_k}} \right)$ denotes the array response vector and is similarly given by \eqref{arrayResponseVectorkl}. For the far-field scenario, the array response vector is reduced to
 \begin{equation}\label{farFieldLocalizationChannel}
 {\bf{a}}\left( {{\bf{w}}; \theta_k }\right) = \left[ e^{ - {\rm{j}}\frac{{2\pi }}{\lambda }y_1\sin \theta_k}, \cdots ,  e^{ - {\rm{j}}\frac{{2\pi }}{\lambda }y_N\sin \theta_k } \right]^T,
 \end{equation}
 where $\theta_k$ denotes the AoA of localization UE $k$, and ${y_n}$ denotes the $y$-coordinate value of ${\bf w}_n$. Let $J$ denote the number of snapshots, the received localization signal at the $j$-th snapshot can be compactly written as
 \begin{equation}\label{localizationReceivedSignal}
 {{\bf{y}}_j}\left( {\bf{w}} \right) = {\bf{A}}\left( {\bf{w}} \right){{\bf{e}}_j} + {{\bf{n}}_j},
 \end{equation}
 where ${\bf{A}}\left( {\bf{w}} \right) \in {{\mathbb C}^{N \times K}} \triangleq \left[ {{\bf{a}}\left( {{\bf{w}};{\theta _1}} \right), \cdots ,{\bf{a}}\left( {{\bf{w}};{\theta _K}} \right)} \right]$, ${{\bf{e}}_j} \in {{\mathbb C}^{K \times 1}} \triangleq {\left[ {{e_{j,1}}, \cdots ,{e_{j,K}}} \right]^T}$, with ${{e_{j,k}}}$ being the independent localization signal of UE $k$ at snapshot $j$, and ${\bf{n}}_j$ denotes the AWGN noise.

 In the following, the Bartlett algorithm is applied to AoA estimation, and its estimation performance is dependent on the array architecture. Specifically, the covariance matrix of the received signal in \eqref{localizationReceivedSignal} is
 \begin{equation}
 {\bf{R}}\left( {\bf{w}} \right) = {\mathbb E}\left[ {{{\bf{y}}_j}\left( {\bf{w}} \right){\bf{y}}_j^H\left( {\bf{w}} \right)} \right] = {\bf{A}}\left( {\bf{w}} \right){{\bf{R}}_e}{{\bf{A}}^H}\left( {\bf{w}} \right) + {\sigma ^2}{{\bf{I}}_N},
 \end{equation}
 where ${{\bf{R}}_e} = {\rm diag}\left( {\left[ {{\Upsilon_1}, \cdots ,{\Upsilon_K}} \right]} \right)$ denotes the covariance matrix of signal ${{\bf{e}}_j}$, with $\Upsilon_{k}$ being the power of signal from localization UE $k$. By averaging all snapshot samples, ${\bf{R}}\left( {\bf{w}} \right)$ is approximately given by
 \begin{equation}
 {\bf{R}}\left( {\bf{w}} \right) = \frac{1}{J}\sum\limits_{j = 1}^J {{{\bf{y}}_j}\left( {\bf{w}} \right){\bf{y}}_j^H\left( {\bf{w}} \right)}.
 \end{equation}

 Furthermore, by vectorizing the matrix ${\bf{R}}\left( {\bf{w}} \right)$, we have
 \begin{equation}\label{covarianceMatrixVec}
 {\bf{z}}\left( {\bf{w}} \right) = {\rm{vec}}\left( {{\bf{R}}\left( {\bf{w}} \right)} \right) = {{\bf{A}}_v}\left( {\bf{w}} \right){\bf{r}} + {\sigma ^2}\overrightarrow {\bf{1}},
 \end{equation}
 where ${{\bf{A}}_v}\left( {\bf{w}} \right)  =\mathbf{A}^*\left( {{\bf{w}}}\right) \odot \mathbf{A}\left( {{\bf{w}}}\right)$, ${\bf{r}} = {\left[ {{\Lambda _1}, \cdots ,{\Lambda _K}} \right]^T}$, and $\overrightarrow {\bf{1}}  = {\left[ {{\bf{f}}_1^T, \cdots ,{\bf{f}}_N^T} \right]^T}$, and ${{\bf{f}}_n}$ denotes a vector where the $n$-th element is  one while all other elements are zero. A closer look at \eqref{covarianceMatrixVec} shows that an equivalent localization signal model is obtained, where ${\bf{A}}_v\left( {\bf{w}} \right)$, ${\bf{r}}$, $\sigma^2\overrightarrow{{\bf{1}}}$ can be viewed as the effective steering matrix, source signal, and noise, respectively. Moreover, after some manipulations, the steering matrix ${\bf{A}}_v\left( {\bf{w}} \right)$ can be written as
 \begin{equation}
 {{\bf{A}}_v}\left( {\bf{w}} \right) = \left[ {{{\bf{a}}_v}\left( {{{\bf{w}}_{\bf{v}}};{\theta _1}} \right), \ldots ,{{\bf{a}}_v}\left( {{{\bf{w}}_{\bf{v}}};{\theta _K}} \right)} \right],
 \end{equation}
 where ${{\bf{a}}_v}\left( {{{\bf{w}}_{\bf{v}}};{\theta _k}} \right) = {{\bf{a}}^*}\left( {{\bf{w}};{\theta _k}} \right) \otimes {\bf{a}}\left( {{\bf{w}};{\theta _k}} \right)$. Note that the conjugate and Kronecker product operations behave like a longer virtual array whose antenna pixels are located at ${\bf{w}}_v$, where ${\bf{w}}_v$ is composed of the distinct values of self/mutual differences of elements within ${\bf{w}}$, known as difference co-array \cite{pal2010nested}. Compared to the $N$-pixel physical array whose array aperture is $\mathcal{O}(N)$, the difference co-array can provide a virtual array aperture of $\mathcal{O}(N^2)$. Moreover, by applying the spatial smoothing, the rank of the covariance matrix of ${\bf{z}}\left( {\bf{w}} \right)$ can be enhanced. 



 With the search vector ${\bf{a}}\left( {{\bf{w}}; \theta }\right)$, where $\theta$ denotes the search angle, the Bartlett spatial spectrum is expressed as
 \begin{equation}
 P_{\rm{Bartlett}}\left({{\bf{w}}; \theta}\right) = {\bf{a}}^H\left( {{\bf{w}}; \theta }\right){\bf{R}}_z{\bf{a}}\left( {{\bf{w}}; \theta }\right).
 \end{equation}
 Then, the estimated AoAs ${{{\hat \theta }_k}\left( {\bf{w}} \right)}$ correspond to the peaks of $P_{\rm{Bartlett}}\left({{\bf{w}}; \theta}\right)$ within the angular search range.

 For localization scenario, the RMSE is adopted as the utility function, given by
 \begin{equation}
 {\rm{RMSE}}\left( {\bf{w}} \right) = \sqrt {\frac{1}{K}\sum\limits_{k = 1}^K {{{\left( {{\theta _k} - {{\hat \theta }_k}\left( {\bf{w}} \right)} \right)}^2}} }.
 \end{equation}

 The array configuration training aims to minimize the RMSE of AoA estimation, and problem \eqref{optimizationProblemACC} becomes
 \begin{equation}\label{arrayTrainingLocalizationProblem}
 \begin{aligned}
 \mathop {\min }\limits_{{\bf w}  }&\ \  {\rm{RMSE}}\left( {\bf{w}} \right)\\
 {\rm{s.t.}}&\ \  {\bf w} \in {\cal W}^{\rm ACC}.
 \end{aligned}
 \end{equation}

 Similarly, exhaustive scanning can be applied to find the best codeword to minimize the RMSE of AoA estimation. Moreover, the proposed two-stage scanning scheme in Section~\ref{subsectionTwoStageScanning} can be utilized to reduce the training overhead. The details are omitted for brevity.

\section{Simulation Results}\label{sectionNumericalResults}
 In this section, simulation results are provided to evaluate the effectiveness of the proposed array configuration training schemes. The carrier frequency and system bandwidth are ${f_c} = 3.5$ GHz and $B = 10$ MHz, respectively. The BS is equipped with an XL-ULA with $M = 256$ antenna pixels, and the number of RF chains is $N = 16$.

  \begin{figure*}[!t]
  \centering
  \subfigure[CA]{
  \begin{minipage}[t]{0.30\linewidth}
    \includegraphics[width=\linewidth]{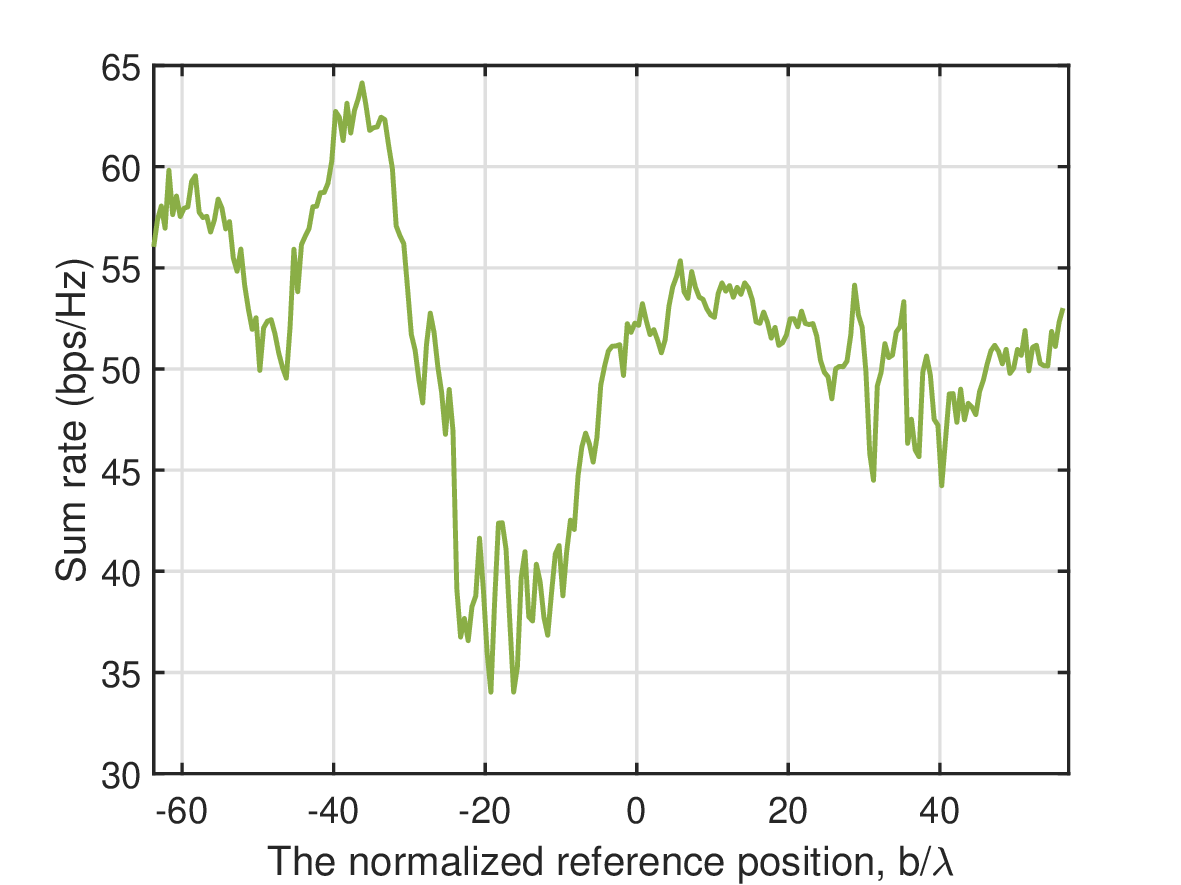}
  \end{minipage}}
  \subfigure[USA]{
  \begin{minipage}[t]{0.30\linewidth}
    \includegraphics[width=\linewidth]{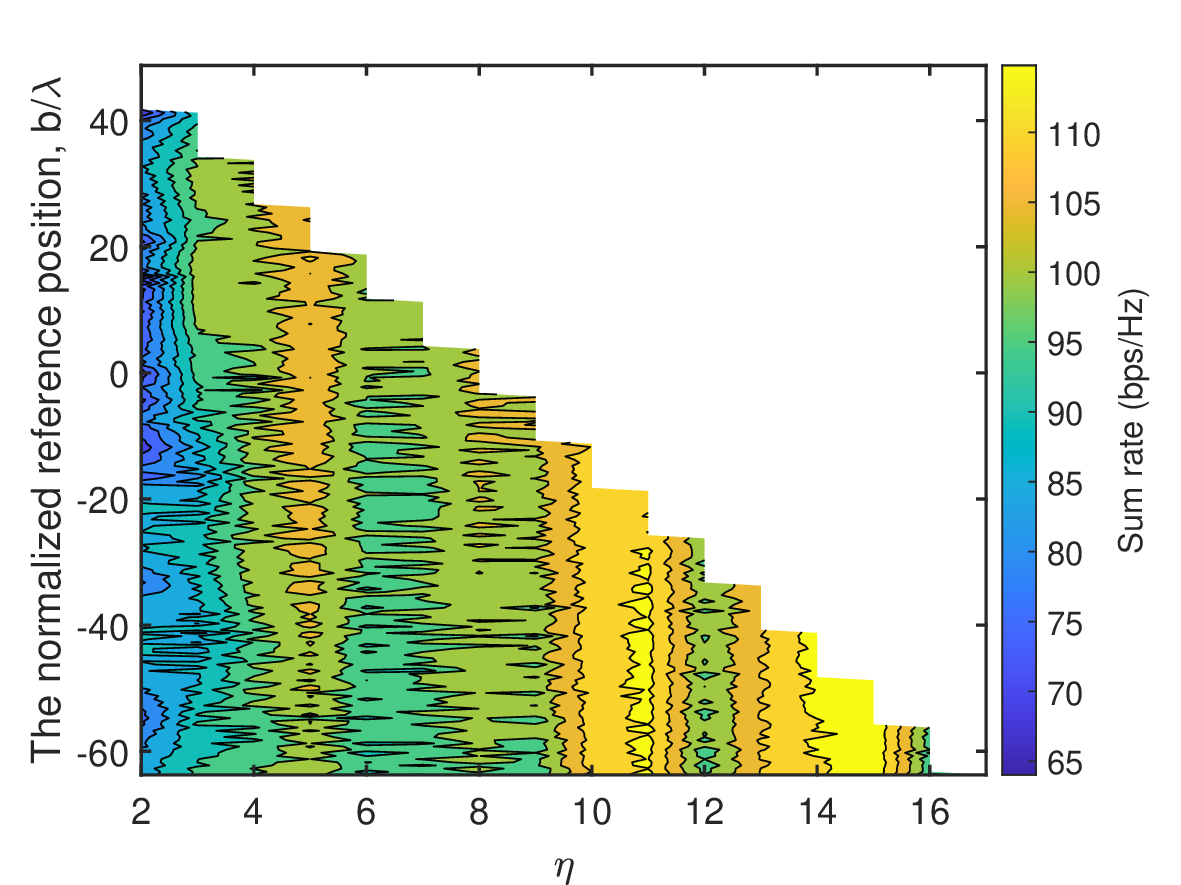}
  \end{minipage}}
  \subfigure[MoA]{
  \begin{minipage}[t]{0.30\linewidth}
    \includegraphics[width=\linewidth]{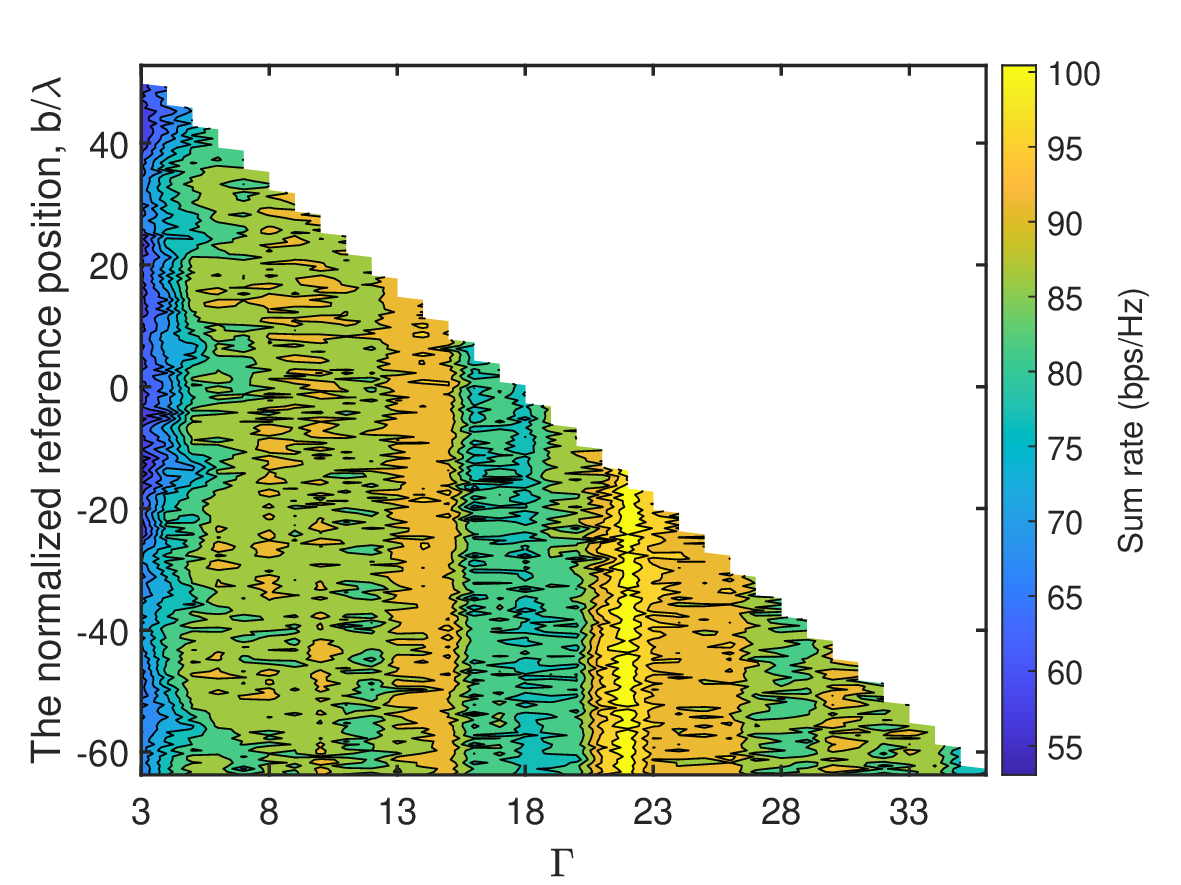}
  \end{minipage}}
  \subfigure[NA]{
  \begin{minipage}[t]{0.30\linewidth}
    \includegraphics[width=\linewidth]{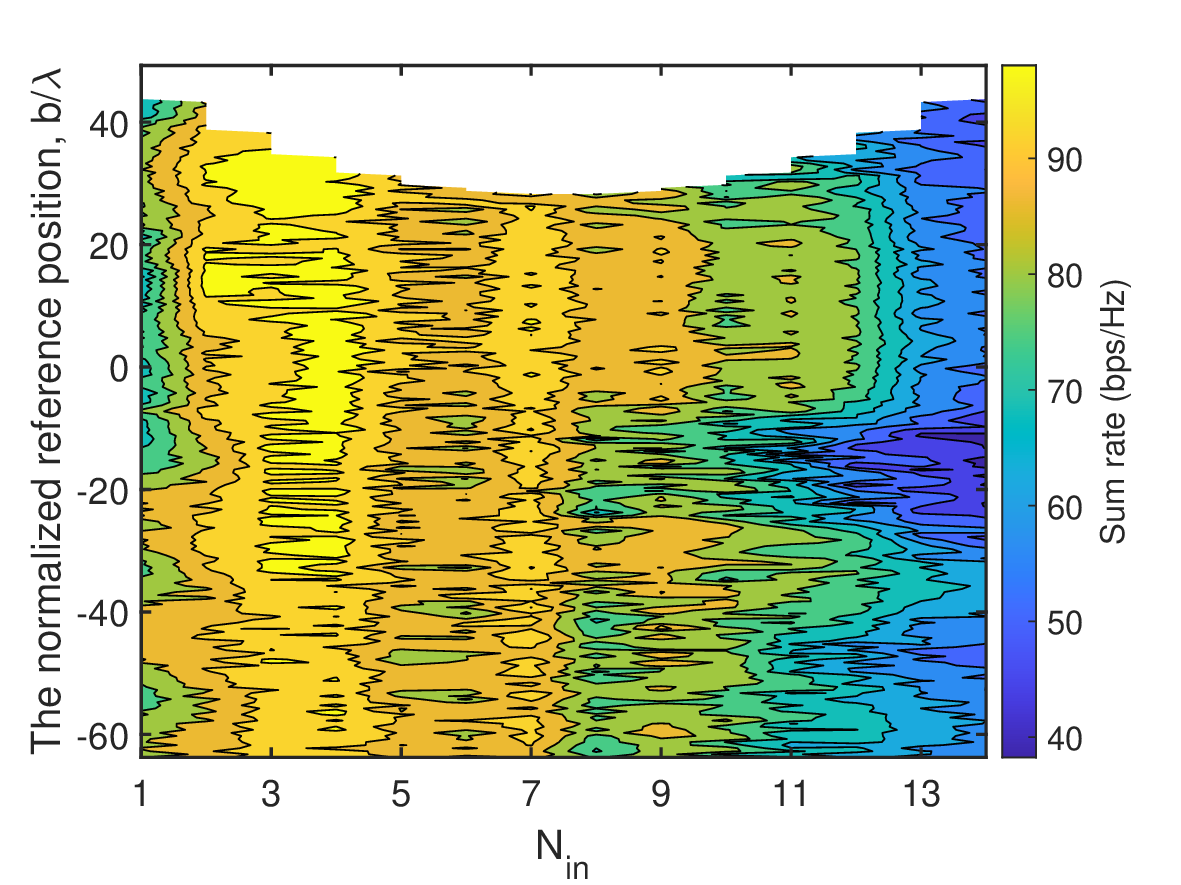}
  \end{minipage}}
  \subfigure[CPA]{
    \begin{minipage}[t]{0.30\linewidth}
    \includegraphics[width=\linewidth]{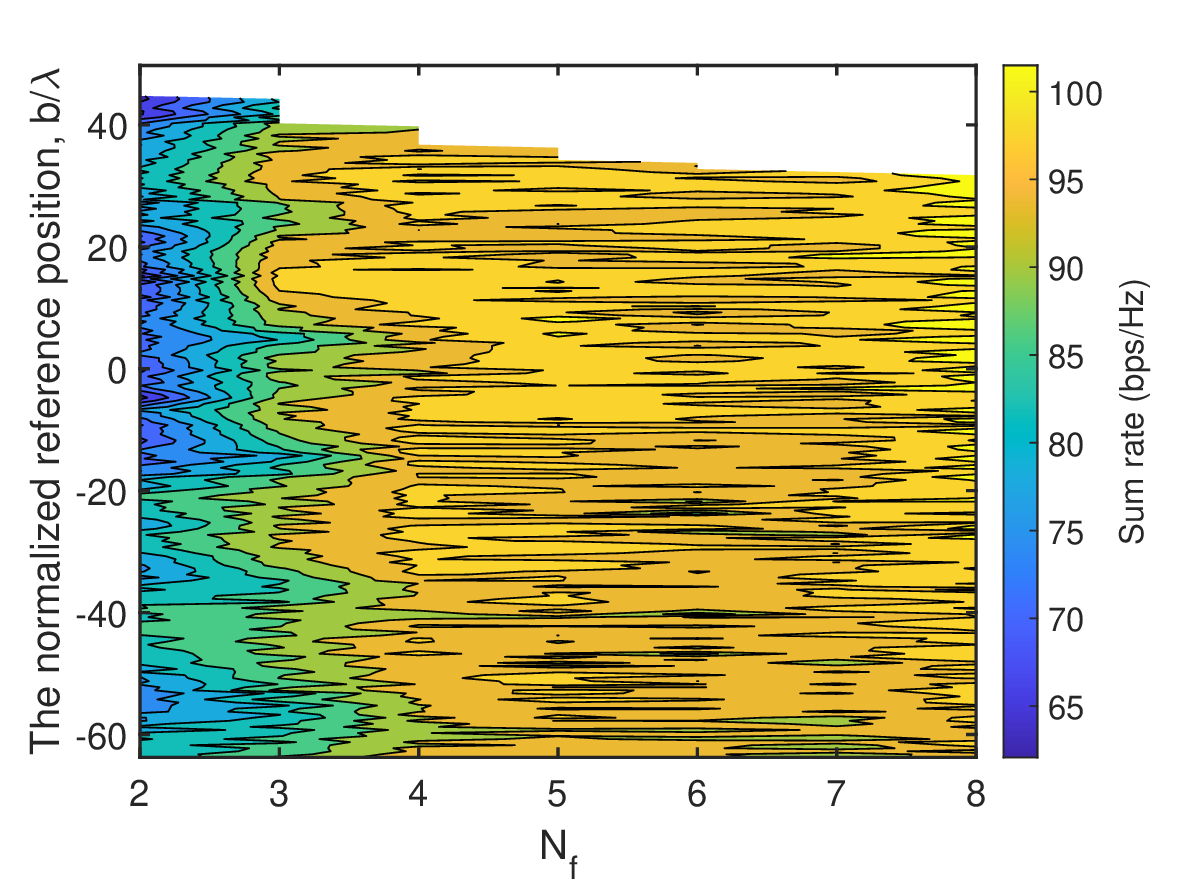}
  \end{minipage}}
  \caption{Sum rate versus array configuration parameter for different array architectures.}
  \label{fig:sumRateVersusParameters}
\end{figure*}

 \subsection{Multi-UE Communication}
 For multi-UE communication, $K$ UEs are uniformly distributed in a circular area specified by the center ${\bf q}_c = {\left[ {200,0} \right]^T}$ m and radius $r_{\rm radius} = 50$ m. The number of scatterers for each UE is ${L_k} =3$, which is uniformly distributed in the area ${r_{k,l}} \in \left( {0,200} \right]$ m, ${\theta _{k,l}} \in \left[ { - {{90}^ \circ },{{90}^ \circ }} \right]$. 
 Unless otherwise stated, the number of UEs is $K = 10$, with the transmit power of UE $k$ being $P_k = 10$ dBm, $1 \le k \le K$.

 Fig.~\ref{fig:sumRateVersusParameters} shows the sum rate versus array configuration parameter for different array architectures, where each array configuration parameter corresponds to one codeword. In contrast to USA, MoA, NA, and CPA, the classic CA can only adjust the reference point position, thus yielding a one-dimensional result shown in Fig.~\ref{fig:sumRateVersusParameters}(a), where the reference position adjustment brings a significant performance variation, with the performance gap between the maximum and minimum sum rates reaching up to $30$ bps/Hz. It is also observed from Figs.~\ref{fig:sumRateVersusParameters}(b)-(e) that the sum rate versus array configuration parameter exhibit different shapes across different array architectures, since the feasible region of each array architecture's reference point position varies with another array configuration parameter. Specifically, for USA, the maximum reference point position decreases as the sparsity level $\eta$ increases, and thus the feasible region of $\eta$ and $b/\lambda$ exhibits a lower triangular shape, and a similar shape can be observed for MoA. Moreover, governed by the array configuration parameters ${N_{\rm in}}$ and ${N_{\rm f}}$, the feasible regions of reference point positions vary for NA and CPA. From Fig.~\ref{fig:sumRateVersusParameters}(b), we observe that for USA, the adjustment of both sparsity level $\eta$ and normalized reference position $b/\lambda$ bring substantial sum rate variations, where its disparity exceeds $55$ bps/Hz in this example. Moreover, similar significant variations in sum rate can be observed for MoA, NA, and CPA in Figs.~\ref{fig:sumRateVersusParameters}(c)-(e), by adjusting the corresponding array configuration parameters. These results demonstrate the necessity of considering ACC, where different codewords yield significant performance disparities.

 \begin{figure}
 \centering
 \subfigure[$K =10$]{
 \begin{minipage}[t]{0.5\textwidth}
 \centering
 \centerline{\includegraphics[width=2.66in,height=2.0in]{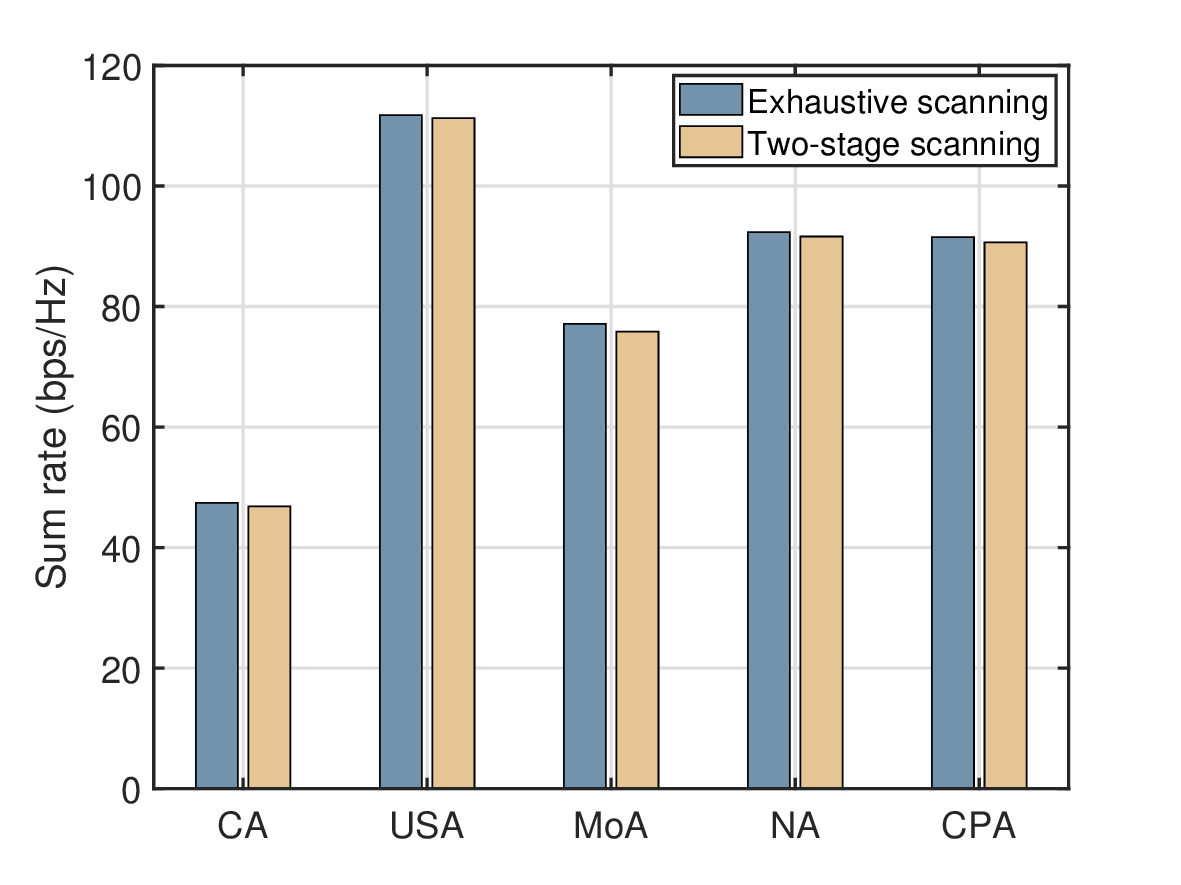}}
 \end{minipage}
 }
 \subfigure[$K =100$]{
 \begin{minipage}[t]{0.5\textwidth}
 \centering
 \centerline{\includegraphics[width=2.66in,height=2.0in]{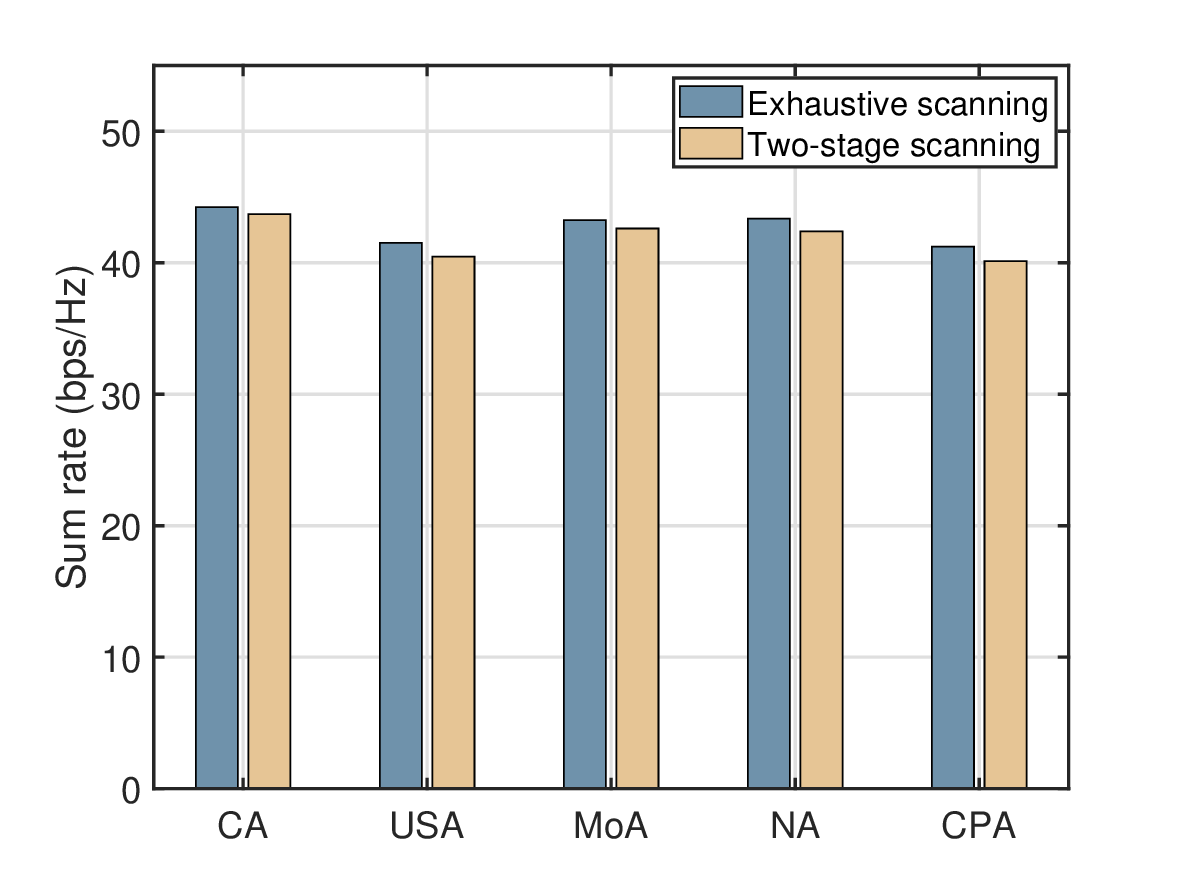}}
 \end{minipage}
 }
 \caption{Comparison of exhaustive and two-stage scanning schemes over different array architectures.}
 \label{exhaustiveVersusTwoStageScanning}
 \end{figure}

 Fig.~\ref{exhaustiveVersusTwoStageScanning} compares the exhaustive and two-stage array configuration scanning schemes over different array architectures for $K = 10$ and $K = 100$, respectively. It is observed that the proposed two-stage scanning scheme yields comparable performance to the exhaustive counterpart for all array architectures, which demonstrates the effectiveness of the proposed two-scanning scheme. It is also observed that the sum rate exhibits significant variations across different array architectures. For example, when the number of UEs is relatively small, e.g., $K=10$, USA gives the best performance. This is because USA can provide the highest spatial resolution to distinguish UEs, thus mitigating the inter-user interference (IUI). However, as the number of UEs increases to $K = 100$, say, the scenario with densely distributed UEs, the classic CA is slightly superior to other four array architectures. This is expected since the other four array architectures suffer from the severe grating lobe issue, particularly for USA where the intensity of grating lobe is identical to that of main lobe \cite{wang2023can,wang2024enhancing}. While for MoA, NA, and CPA, the intensity of grating lobe is smaller than that of main lobe \cite{li2024multi,min2025integrated}. The above results verify the necessity for considering distinct array architectures for different numbers of UEs.

 \begin{figure}[!t]
 \centering
 \centerline{\includegraphics[width=3.0in,height=2.25in]{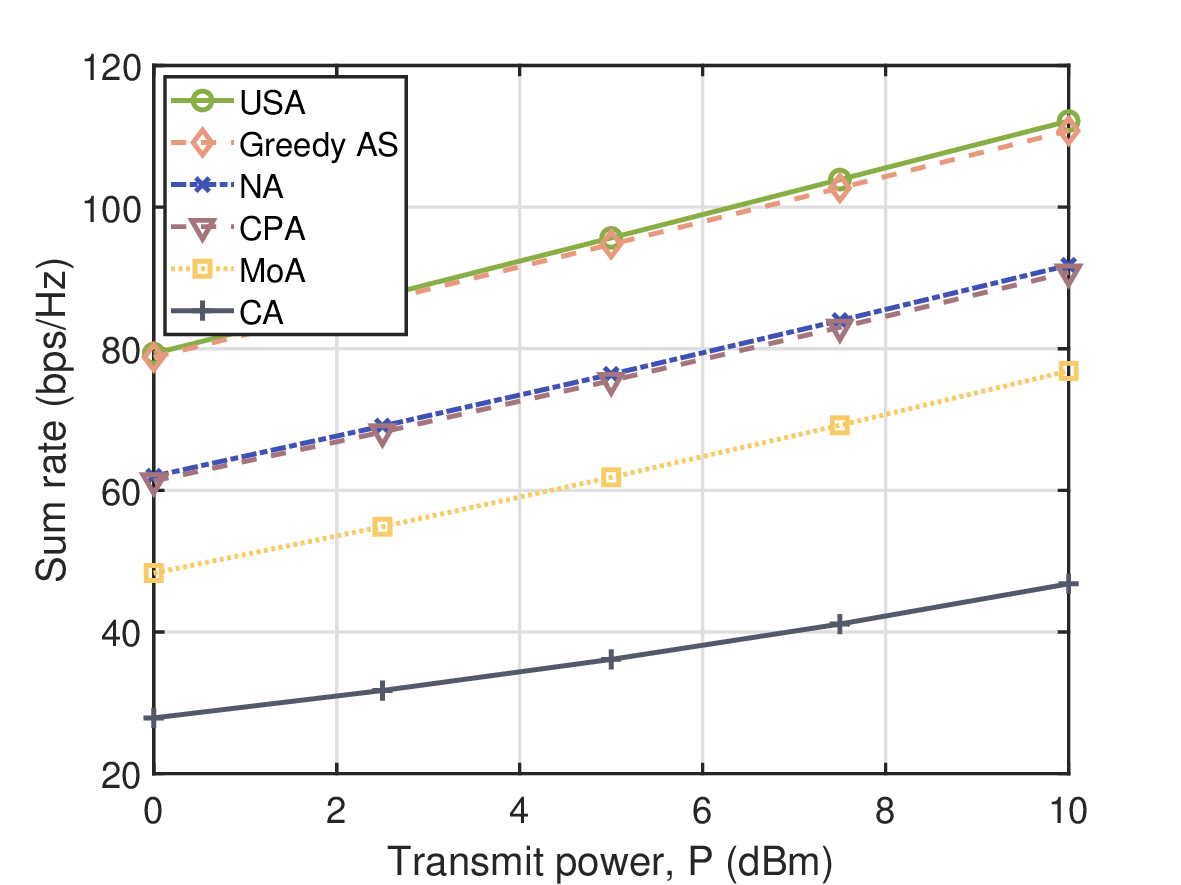}}
 \caption{Sum rate versus the transmit power of each UE.}
 \label{fig:sumRateVersusTransmitPower}
 \end{figure}

 Fig.~\ref{fig:sumRateVersusTransmitPower} shows the sum rate versus the transmit power of each UE for greedy AS and two-stage scanning scheme over different array architectures. It is observed that the two-stage scanning scheme over USA gives comparable performance to the greedy AS scheme, and they are superior to the two-stage scanning schemes over other four array architectures. Meanwhile, the worst-case training overhead of the two-stage scanning scheme over USA is $| {{\cal W}_{\rm USA}^{\left( 1 \right)}} | + (M-1)/2 = 35 + 127 = 162$, while the number of computations for greedy AS is $\left( {2M - N + 1} \right)N/2 = 3976$. This demonstrates the superiority of the proposed two-stage scanning scheme over greedy AS in terms of training overhead reduction.

 \begin{figure}[!t]
 \centering
 \centerline{\includegraphics[width=3.0in,height=2.25in]{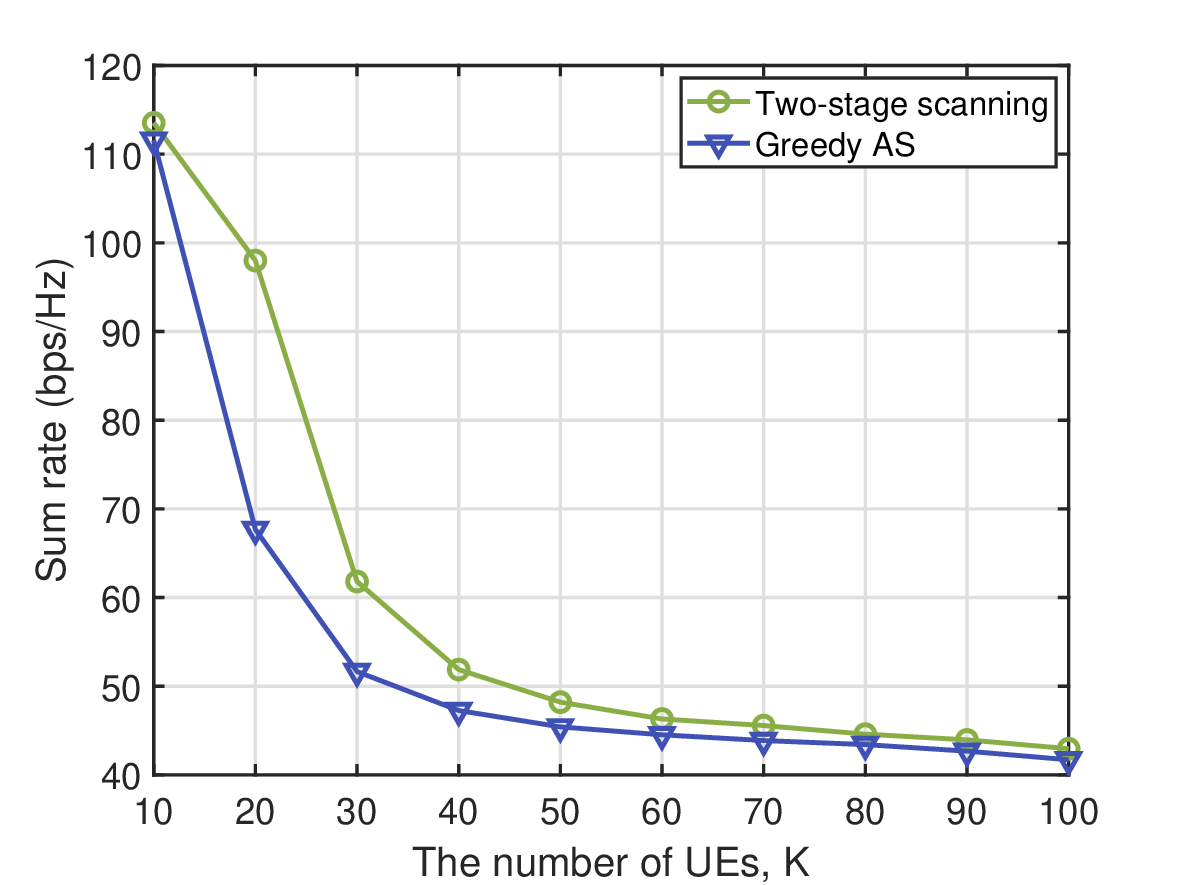}}
 \caption{Sum rate versus the number of UEs.}
 \label{fig:sumRateVersusUENumber}
 \end{figure}

 Furthermore, Fig.~\ref{fig:sumRateVersusUENumber} shows the sum rate versus the number of UEs, $K$. It is observed that the performance of both the greedy AS and the two-stage scanning scheme over ACC exhibit a decline trend as $K$ increases. This is expected since an increase in the number of UEs results in a more severe IUI issue. It is also observed that the performance gap between the two-stage scanning scheme over ACC and greedy AS schemes first increases and finally diminishes. This is because when the number of UEs is relatively small, the two-stage scanning scheme over ACC is able to obtain an appropriate array architecture with optimized array configuration parameter for mitigating the IUI. However, as $K$ further increases, the ACC enabled flexible array architecture cannot deal with the severe IUI issue.

 \subsection{Wireless Localization}
 For the wireless localization scenario, $K$ localization UE signals with equal power are uniformly distributed in $\left[ { - {{45}^ \circ },{{45}^ \circ }} \right]$. The number of snapshots is $J = 1000$, and the two-stage scanning scheme over each array architecture is applied to obtain the corresponding array configuration parameter.

 \begin{figure}[!t]
 \centering
 \centerline{\includegraphics[width=3.0in,height=2.25in]{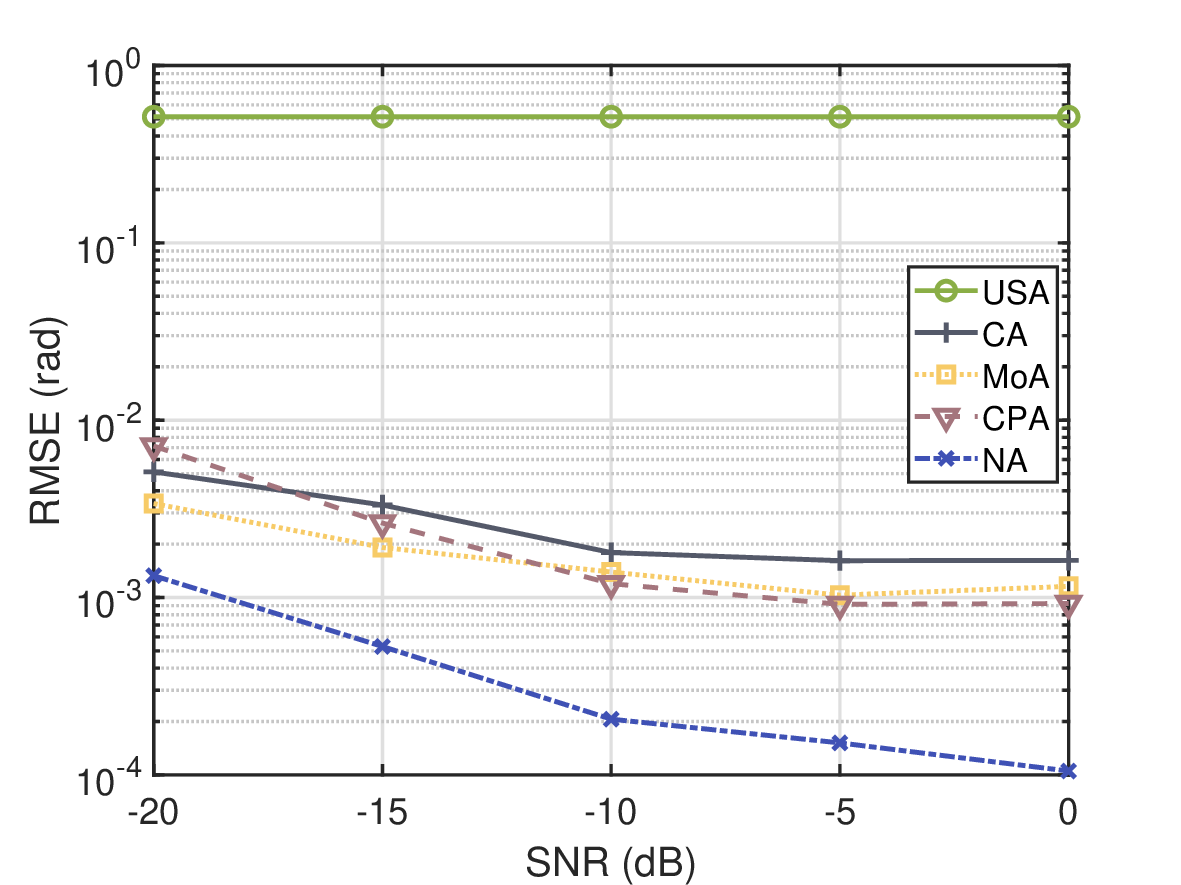}}
 \caption{Localization RMSE versus the SNR.}
 \label{fig:RMSEVersusSNR}
 \end{figure}

 Fig.~\ref{fig:RMSEVersusSNR} shows the localization RMSE versus the SNR. The SNR is defined as the ratio of signal power to noise power, and the number of localization UEs is $K = 8$. It is firstly observed that the localization RMSE of USA remains almost unchanged as the SNR increases, which is mainly attributed to the angular ambiguity issue caused by grating lobe, rendering it challenging for USA to estimate the AoA, even with the array configuration parameter adjustment. This stands in stark contrast to a superior sum rate performance achieved by USA in multi-UE communications (as can be seen in Fig.~\ref{exhaustiveVersusTwoStageScanning}(a)). It is also observed that the localization RMSE of the other four array architectures exhibit a decreasing trend as the SNR increases, as expected. Moreover, the RMSE performance of non-uniform sparse arrays, e.g., MoA, NA, and CPA, in general outperforms the classic compact array, and NA yields a best RMSE performance among the five array architectures. This is because NA is able to provide the largest continuous virtual aperture by utilizing the difference co-array.

 \begin{figure}[!t]
 \centering
 \centerline{\includegraphics[width=3.0in,height=2.25in]{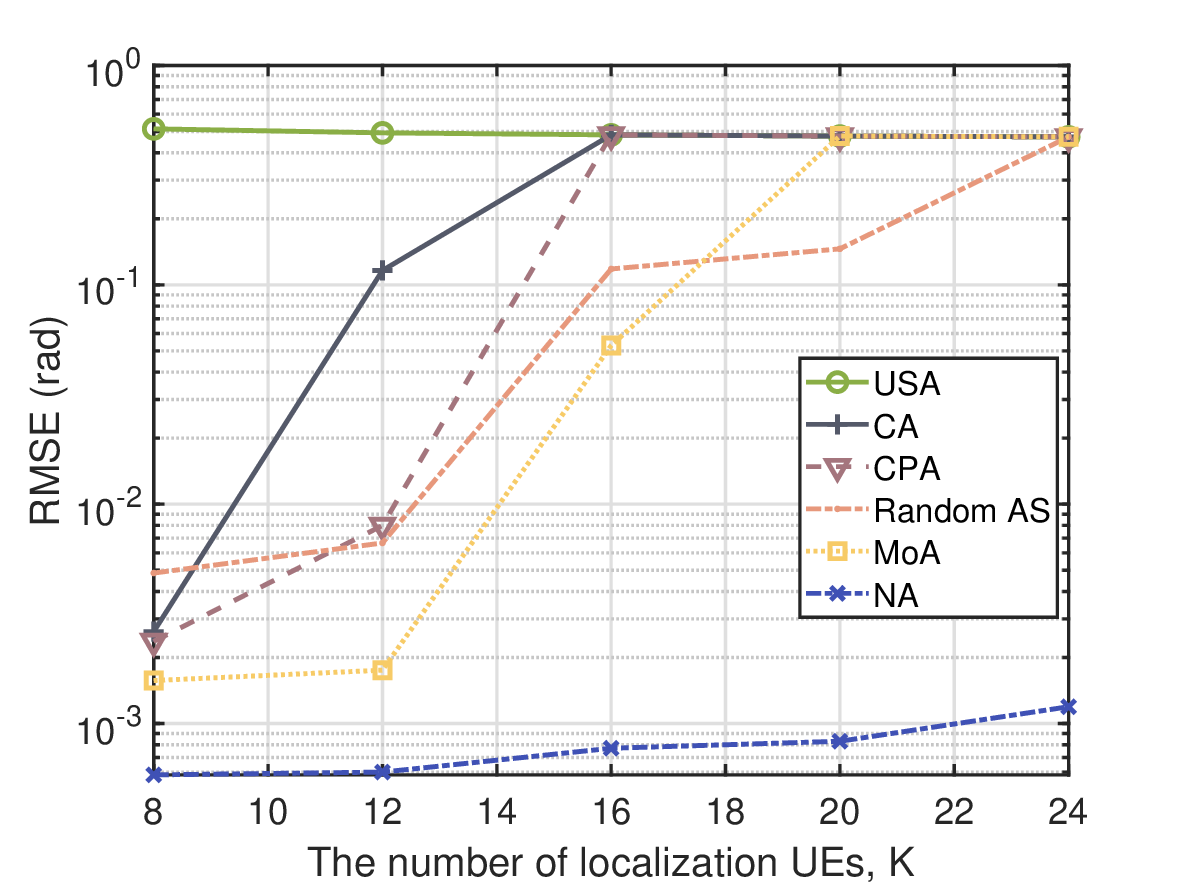}}
 \caption{Localization RMSE versus the number of localization UEs.}
 \label{fig:RMSEVersusLocalizationUENumber}
 \end{figure}

 Fig.~\ref{fig:RMSEVersusLocalizationUENumber} shows the localization RMSE versus the number of localization UEs, $K$. The SNR is set as $-15$ dB. For comparison, the random AS scheme is considered, where the activated antenna pixel positions are randomly generated, and the RMSE is obtained by selecting the one with the minimum RMSE from $10^4$ independent realizations. It is observed that the localization RMSE performance of all schemes deteriorate as $K$ increases, which is expected since more sources need to be estimated. Moreover, when the number of localization UEs is larger than that of antenna pixels, i.e., $K \ge 16$, the classic CA with the array configuration adjustment fails to estimate the AoAs, which arises from the fact that it can only provide the localization DoF no greater than the number of antenna pixels. By contrast, NA is capable of estimating more number of localization UEs without compromising too much estimation accuracy, thanks to the enlarged DoF brought by the increased virtual aperture and flexible array configuration parameter adjustment.

 Last, by comparing the results of multi-UE communication and wireless localization, it can be found that it is preferable to adopt different array architectures for different scenarios. For example, the combinations of CA and USA can be utilized for communication scenario, while non-uniform sparse arrays, such as MoA and NA, can be utilized for localization scenario, which thus reduces the codewords to be scanned and training overhead. Furthermore, since the two-stage scanning scheme can be performed in a parallel manner, and dynamic pixel activation can be efficiently achieved via electronic control with low latency, it is possible to reconfigure the XL-MIMO architecture in a small timescale.

\section{Conclusion}\label{sectionConclusion}
 This paper proposed a novel concept called ACC, which enables the cost-effective implementation of XL-MIMO. We presented an ACC design example, which is composed of the CA, USA, as well as NUSA encompassing MoA, NA, and CPA. Specifically, the codewords for each array architecture vary with array configuration parameters, where the reference position is the common array configuration parameter, while sparsity level, inter-module spacing level, and the number of inner/first array's antenna pixels constitute architecture-specific parameters for USA, MoA, and NA/CPA, respectively. With the designed ACC, we developed an efficient two-stage array configuration training scheme for multi-UE communication, so as to reduce the training overhead. As a comparison, a greedy AS scheme tailored for multi-UE communication was proposed. Then, the two-stage scanning scheme was extended to the wireless localization scenario. Simulation results verified the significant performance variations across different codewords, and the necessity of adopting different array architectures for different scenarios. 
 
 In the future, there are some promising research directions. For example, extending the proposed ACC design to encompass additional array architectures, e.g., MRA and MHA, is worthwhile to investigate for further enhancing the system performance. Moreover, the extensions of the proposed scheme to stochastic channel models and downlink communication scenarios require more in-depth studies.

\begin{appendices}
\section{Proof of Proposition \ref{innerArrayElementNumberProposition}}\label{proofinnerArrayElementNumberProposition}
 To form the NA, we have ${N_{{\rm{in}}}} \ge 1$ and ${N_{{\rm{ou}}}} = N - {N_{{\rm{in}}}} \ge 2$, yielding ${N_{{\rm{in}}}} \in \left\{ {1,2, \cdots ,N - 2} \right\}$. Moreover, by substituting ${N_{{\rm{ou}}}} = N - {N_{{\rm{in}}}}$ into \eqref{dimensionConstraintNA} and after some manipulations, it follows that
 \begin{equation}\label{dimensionConstraintNA2}
 N_{{\rm{in}}}^2 - \left( {N - 1} \right){N_{{\rm{in}}}} + M - N \ge 0.
 \end{equation}
 By checking the inequality, we have the following two cases.

 \emph{Case 1:} ${\left( {N + 1} \right)^2}/4 \le M$. In this case, \eqref{dimensionConstraintNA2} always holds. However, to form NA, the number of inner antenna pixels should satisfy ${N_{{\rm{in}}}} \in \left\{ {1,2, \cdots ,N - 2} \right\}$.

 \emph{Case 2:} ${\left( {N + 1} \right)^2}/4 > M$. In this case, the solution to \eqref{dimensionConstraintNA2} is ${N_{\rm{in}}} \le \frac{{\left( {N - 1} \right) - \sqrt {{{\left( {N + 1} \right)}^2} - 4M} }}{2}$ or ${N_{\rm{in}}} \ge \frac{{\left( {N - 1} \right) + \sqrt {{{\left( {N + 1} \right)}^2} - 4M} }}{2}$. Since the number of inner antenna pixels is an integer, we have ${N_{{\rm{in}}}} \le N_{{\rm{in}}}^l$ or ${N_{{\rm{in}}}} \ge N_{{\rm{in}}}^u$, where $N_{{\rm{in}}}^l$ and $N_{{\rm{in}}}^u$ are defined below \eqref{feasibleRegionNin}, respectively. Depending on the relationship between $N_{{\rm{in}}}^l$ and 1, the following two subcases are involved.

 \emph{Subcase 1:} $N_{{\rm{in}}}^l < 1$. It can be verified that $N_{{\rm{in}}}^u = N - 1$. Thus, in this case, it is unable to form  NA.

 \emph{Subcase 2:} $N_{{\rm{in}}}^l \ge 1$. In this case, the feasible region of ${N_{{\rm{in}}}}$ is $\left\{ {1, \cdots ,N_{{\rm{in}}}^l} \right\} \cup \left\{ {N_{{\rm{in}}}^u, \cdots ,N - 2} \right\}$.

 By integrating the above two cases, the proof of Proposition \ref{innerArrayElementNumberProposition} is thus completed.

\section{Proof of Proposition \ref{incrementalExpressionofSINRProposition}}\label{proofIncrementalExpressionofSINRProposition}

 When antenna pixel $m$ is selected at the $n$-th step, and with the definition of ${{\bf{H}}_k}$ above \eqref{covarianceMatrix}, we have ${\bf{H}}_k^{\left( n \right)} = {[ {{{( {{\bf{H}}_k^{\left( {n - 1} \right)}} )^T}},{{( {{\bf{d}}_{k,m}^H} )^T}}} ]^T}$. It then follows that
 \begin{equation}\label{HkTranspositionHk}
 {\left( {{\bf{H}}_k^{\left( n \right)}} \right)^H}{\bf{H}}_k^{\left( n \right)}= {\left( {{\bf{H}}_k^{\left( {n - 1} \right)}} \right)^H}{\bf{H}}_k^{\left( {n - 1} \right)} + {{\bf{d}}_{k,m}}{\bf{d}}_{k,m}^H.
 \end{equation}
 By substituting \eqref{HkTranspositionHk} into  ${{\bf{G}}_k^{\left( n \right)}}$, we have
 \begin{equation}
 {\bf{G}}_k^{\left( n \right)} = {\bf{G}}_k^{\left( {n - 1} \right)} - \frac{{{\bf{G}}_k^{\left( {n - 1} \right)}{{\bf{d}}_{k,m}}{\bf{d}}_{k,m}^H{\bf{G}}_k^{\left( {n - 1} \right)}}}{{1 + {\bf{d}}_{k,m}^H{\bf{G}}_k^{\left( {n - 1} \right)}{{\bf{d}}_{k,m}}}}.
 \end{equation}
 Based on \eqref{alternativeReceivedSINRUEk}, the resulting SINR for UE $k$ can be expressed as \eqref{incrementalExpressionofSINRProof}, shown at the top of the next page. The proof of Proposition \ref{incrementalExpressionofSINRProposition} is thus completed.

 \newcounter{mytempeqncnt1}
 \begin{figure*}
 \normalsize
 \setcounter{mytempeqncnt1}{\value{equation}}
 \begin{equation}\label{incrementalExpressionofSINRProof}
 \begin{aligned}
 \gamma _{k,m}^{\left( n \right)} &= {{\bar P}_k}\left( {{{\left\| {{\bf{\tilde h}}_k^{\left( n \right)}} \right\|}^2} - {{\left( {{\bf{\tilde h}}_k^{\left( n \right)}} \right)}^H}{\bf{H}}_k^{\left( n \right)}{\bf{G}}_k^{\left( n \right)}{{\left( {{\bf{H}}_k^{\left( n \right)}} \right)}^H}{\bf{\tilde h}}_k^{\left( n \right)}} \right) = {{\bar P}_k}\left( {{{\left\| {{\bf{\tilde h}}_k^{\left( {n - 1} \right)}} \right\|}^2} + {{\left| {{h_{k,m}}} \right|}^2} - } \right.\\
 &\ \ \ \left. {\left[ {{{\left( {{\bf{\tilde h}}_k^{\left( {n - 1} \right)}} \right)}^H},h_{k,m}^*} \right]\left[ \begin{split}
 &{\bf{H}}_k^{\left( {n - 1} \right)}\\
 &{\bf{d}}_{k,m}^H
 \end{split} \right]\left( {{\bf{G}}_k^{\left( {n - 1} \right)} - \frac{{{\bf{G}}_k^{\left( {n - 1} \right)}{{\bf{d}}_{k,m}}{\bf{d}}_{k,m}^H{\bf{G}}_k^{\left( {n - 1} \right)}}}{{1 + {\bf{d}}_{k,m}^H{\bf{G}}_k^{\left( {n - 1} \right)}{{\bf{d}}_{k,m}}}}} \right)\left[ {{{\left( {{\bf{H}}_k^{\left( {n - 1} \right)}} \right)}^H},{{\bf{d}}_{k,m}}} \right]\left[ \begin{split}
 &{\bf{\tilde h}}_k^{\left( {n - 1} \right)}\\
 &{h_{k,m}}
 \end{split} \right]} \right)\\
 &= {{\bar P}_k}\left( {{{\left\| {{\bf{\tilde h}}_k^{\left( {n - 1} \right)}} \right\|}^2} - {{\left( {{\bf{\tilde h}}_k^{\left( {n - 1} \right)}} \right)}^H}{\bf{H}}_k^{\left( {n - 1} \right)}{\bf{G}}_k^{\left( {n - 1} \right)}{{\left( {{\bf{H}}_k^{\left( {n - 1} \right)}} \right)}^H}{\bf{\tilde h}}_k^{\left( {n - 1} \right)}} \right) + {{\bar P}_k}\left( {\frac{{{{\left| {{h_{k,m}}} \right|}^2}}}{{1 + {\bf{d}}_{k,m}^H{\bf{G}}_k^{\left( {n - 1} \right)}{{\bf{d}}_{k,m}}}} + } \right.\\
 &\ \ \ \Bigg. {\frac{{{{\left| {{{\left( {{\bf{\tilde h}}_k^{\left( {n - 1} \right)}} \right)}^H}{\bf{H}}_k^{\left( {n - 1} \right)}{\bf{G}}_k^{\left( {n - 1} \right)}{{\bf{d}}_{k,m}}} \right|}^2}}}{{1 + {\bf{d}}_{k,m}^H{\bf{G}}_k^{\left( {n - 1} \right)}{{\bf{d}}_{k,m}}}} - h_{k,m}^*\frac{{{\bf{d}}_{k,m}^H{\bf{G}}_k^{\left( {n - 1} \right)}{{\left( {{\bf{H}}_k^{\left( {n - 1} \right)}} \right)}^H}{\bf{\tilde h}}_k^{\left( {n - 1} \right)}}}{{1 + {\bf{d}}_{k,m}^H{\bf{G}}_k^{\left( {n - 1} \right)}{{\bf{d}}_{k,m}}}} - {h_{k,m}}\frac{{{{\left( {{\bf{\tilde h}}_k^{\left( {n - 1} \right)}} \right)}^H}{\bf{H}}_k^{\left( {n - 1} \right)}{\bf{G}}_k^{\left( {n - 1} \right)}{{\bf{d}}_{k,m}}}}{{1 + {\bf{d}}_{k,m}^H{\bf{G}}_k^{\left( {n - 1} \right)}{{\bf{d}}_{k,m}}}}} \Bigg)\\
 & = \gamma _k^{\left( {n - 1} \right)} + {{\bar P}_k}\frac{{{{\left| {h_{k,m}^ *  - {{\left( {{\bf{\tilde h}}_k^{\left( {n - 1} \right)}} \right)}^H}{\bf{H}}_k^{\left( {n - 1} \right)}{\bf{G}}_k^{\left( {n - 1} \right)}{{\bf{d}}_{k,m}}} \right|}^2}}}{{1 + {\bf{d}}_{k,m}^H{\bf{G}}_k^{\left( {n - 1} \right)}{{\bf{d}}_{k,m}}}}.
 \end{aligned}
 \end{equation}
 \hrulefill
 \vspace{-0.5cm}
 \end{figure*}

\end{appendices}

\bibliographystyle{IEEEtran}
\bibliography{refFlexibleAntenna}

\end{document}